\documentclass[preprint,showpacs,preprintnumbers,nofootinbib,amsmath,amssymb]{revtex4}

\usepackage{graphicx}
\usepackage{dcolumn}
\usepackage{bm}
\usepackage{float}

\newcommand{\der}[1]{\frac{\partial}{\partial #1}}
\newcommand{\sss}[1]{\scriptscriptstyle #1}

\newcommand{\bo}[1]{\mbox{\boldmath $ #1 $}}
\newcommand{\s}[1]{{\rlap/ #1}}
\restylefloat{figure}
\restylefloat{table}

\begin{document}

\title{Bootstrap and the physical values of
$\pi N$ resonance parameters.}

\author{K.~Semenov-Tian-Shansky}
\email{cyr_stsh@mail.ru}
\altaffiliation[Also at ]{Institut f\"{u}r Theoretische Physik II,
Ruhr-Universit\"{a}t Bochum, D-44780 Bochum, Germany}
\author{Alexander V.~Vereshagin}%
 \email{Alexander.Vereshagin@ift.uib.no}
\altaffiliation[Also at ]{University of Bergen,
Institute of Physics and Technology,
Allegt.55, N5007 Bergen, Norway }
\author{Vladimir V.~Vereshagin}
\email{vvv@AV2467.spb.edu}
\affiliation{%
St.-Petersburg State University,
St.Petersburg, Petrodvoretz, 198504, Russia
}%

\begin{abstract}
This is the 6th paper in the series developing the formalism to manage
the effective scattering theory of strong interactions. Relying on the
theoretical scheme suggested in our previous publications we
concentrate here on the practical aspect and apply our technique to
the elastic pion-nucleon scattering amplitude. We test numerically the
$\pi N$
spectrum sum rules that follow from the tree level bootstrap
constraints. We show how these constraints can be used to estimate
the tensor and vector
$NN\rho$
coupling constants. At last, we demonstrate that the tree-level low
energy expansion coefficients computed in the framework of our
approach show nice agreement with known experimental data. These
results allow us to claim that the extended
perturbation scheme is quite
reasonable from the computational point of view.

\end{abstract}

\maketitle

\section{Introduction}
\label{sec-Introduction}

In our previous publications (see
\cite{AVVV2,KSAVVV2})
we developed the generic construction of efficient perturbation scheme
intended for effective scattering theories of strong interaction%
\footnote{Preliminary analysis has been published in
\cite{POMI} - \cite{talks}.}.
This study is still in progress. Meanwhile, already our present
results appear to be quite sufficient to justify the usage of
experimental data for checking the correctness of
{\em tree level}
bootstrap constraints for the effective theory parameters.

Due to the renormalization invariance of bootstrap constraints (see
\cite{KSAVVV2})
those constrains of arbitrary loop level present exact
(self-consistency type) numerical limitations for the admissible
values of renormalization prescriptions. These prescriptions are the
only fundamental observables of a theory and, hence, every kind of
theoretical constraints for their values can be directly compared with
experimental data. This is true irrelatively to the loop order of the
bootstrap constraints under consideration. For this reason it seems us
natural to perform the numerical testing of the tree level bootstrap
constrains using the available experimental data. This will allow us
to check the physical reasonability of our main postulates and, at the
same time, to demonstrate the practical output of the formalism
discussed in the above-cited articles.

This paper is designed as a regular introduction to the corresponding
calculational methods. We demonstrate the details of calculational
procedure beginning with general formulae and ending with numerical
results.

As an example we consider below the elastic pion-nucleon scattering
process. We derive and compare with known data several sum rules
for the parameters (coupling constants and masses) of pion-nucleon
resonances that follow from the tree level bootstrap constraints.

Besides, we show that the latter constraints provide reasonable
estimates for the numerical values of experimentally known
(see, e.g.,
\cite{Nagels})
phenomenological constants
$G_{\scriptscriptstyle T}$
and
$G_{\scriptscriptstyle V}$
which describe the tensor and vector types of
$\rho$-meson
coupling to nucleon.

Finally, we present the results for tree level values of low energy
expansion coefficients of pion-nucleon scattering amplitude around the
cross-symmetric point
$(t,\nu_t)=(0,0)$.
The values of these coefficients are, by no doubt, affected by loop
corrections. Nevertheless, as follows from our estimates, the tree
level values obtained in the framework of extended perturbation scheme
turn out to be very close to the experimental ones. This fact suggests
that the extended perturbation scheme is quite reasonable from the
physical point of view.

\section{Preliminaries}
\label{sec-Preliminaries}

In this Section we quote those results of the papers
\cite{AVVV2,KSAVVV2}
which constitute the theoretical background of our
calculations below.
It is implied that the reader is familiar with the
notions and terminology introduced in those articles.

First of all let us remind that we only consider a special class of
effective theories called in
\cite{KSAVVV2}
as
{\em localizable}.
To assign meaning to individual terms of Dyson series for such a
theory we switch to the so-called
{\em extended perturbation scheme}
which contains supplementary resonance fields. This procedure can be
treated as a special kind of summation of an infinite set of graphs
(with the same number of loops) that appear in every order of the
initial Dyson series.

The extended perturbation scheme is just an auxiliary construction
which allows us to define rigorously the perturbation expansion in the
case of infinite component effective theory. In particular, the
$S$-matrix
calculated in the framework of extended perturbation scheme still acts
on the space of asymptotic states that correspond to true stable (with
respect to strong decays) particles. The supplementary resonance
fields do not correspond to any asymptotic states and hence may appear
only in the inner lines of graphs which describe the scattering
processes of stable particles. In this paper we consider the case when
there are only two species of stable particles, namely, pions and
nucleons.

The list of the results of
\cite{AVVV2,KSAVVV2}
which we rely upon in this paper reads:
\begin{itemize}
\item
In the framework of effective theory an arbitrary renormalized
$S$-matrix
graph can be presented in the form solely constructed from the minimal
propagators and resultant vertices of various levels. The true loop
order of a given graph is just a number of explicitly drawn loops plus
the sum of level indices of its vertices.
\item
All the information needed to completely fix the kinematical structure
of renormalized
$S$-matrix
elements of a given loop order
$L$
is contained in the numerical values of resultant parameters of
$L$th
and lower levels.
\item
By construction, the resultant parametrization implies using the
scheme of
{\em renormalized perturbation theory}.
This means that the relevant resultant parameters (in the case we
analyze below --- the 0th level ones) should be considered as
fundamental physical observables of the theory.
\end{itemize}

These results are based on
{\em summability}
and
{\em uniformity}
requirements which are the corner stones of our extended perturbation
scheme. The motivation for accepting these two requirements is
presented in
\cite{KSAVVV2}.

The uniformity requirement is formulated as follows:
{\em the degree of the bounding polynomial which specifies the
asymptotics of a given loop order amplitude must be the same as that
specifying the asymptotics of the full (non-perturbative) amplitude
of the process under consideration.}

The summability requirement reads:
{\em in every sufficiently small domain of the complex space of
kinematical variables there must exist an appropriate order of
summation of the formal series of contributions coming from the
graphs  with given number of loops, such that the reorganized series
converges. Altogether, these series must define a unique analytic
function with only those singularities that are presented in the
contributions of individual graphs.}

As a system of domains in which we require the
$2 \rightarrow 2$
amplitude to be summable we choose three hyperlayers
\begin{equation}
B_x \left\{ x \in { \mathbb{R}},\  {\nu}_x \in { \mathbb{C}};\ \;
x \sim 0 \right\},\ \ \ \ \ \ (x=s,t,u).
\nonumber 
\end{equation}
Here
$s,t,u$
stand for conventional Mandelstam variables; the energy-like variables
$\nu_x$
are defined as follows:
\begin{equation}
{\nu}_s \equiv (u-t);\ \ \ \ \ \
{\nu}_t \equiv (s-u);\ \ \ \ \ \
{\nu}_u \equiv (t-s).
\label{x_nu}
\end{equation}
We imply that the full amplitudes under consideration satisfy Regge
asymptotic conditions, at least, at sufficiently small values of the
momentum transfer. With respect to tree level
\mbox{$2 \rightarrow 2$}
amplitudes this
means that they are described by the polynomially bounded meromorphic
functions of pair energies (at fixed value of the corresponding
momentum transfer). The bounding polynomial degree in every hyperlayer
$B_x$
is fixed by the value of the relevant Regge intercept.

The results of
\cite{KSAVVV2}
define  the sequence of steps one should follow to derive the tree level
bootstrap constraints for
\mbox{$2 \rightarrow 2$}
scattering amplitude:


\begin{enumerate}
\item
Consider the general structure of the amplitude and single out the
invariant formfactors.
\item
Draw all loopless graphs for the amplitude of the process under
consideration using Feynman rules of the extended perturbation scheme.
\item
Classify the possible types of triple vertices in accordance with
quantum numbers of the line corresponding to a virtual particle.
\item
Construct the analytic expressions for individual graph contributions
only using the minimal propagators and resultant vertices.
\item
Perform a
{\em formal}
summation over all possible kinds of vertices and internal lines.
This will result in the formal infinite sum of pole terms coming from
the resonance exchange graphs plus a formal power series in two
independent variables stemming from the pointlike vertices.
\item
Now, being guided by summability and uniformity principles and
applying the technique of Cauchy forms, convert a disordered sum of
amplitude graphs into a rigourously defined expressions in three
hyperlayers
$B_x$ $(x=s,t,u)$.
The principle parts of the corresponding Cauchy forms are
determined by the individual resonance exchange contributions.
The bounding polynomial degrees are dictated by the values of
corresponding Regge intercepts.
\item
In three intersection domains
$$
D_s = B_t \cap B_u,\ \ \ \ \ \
D_t = B_u \cap B_s,\ \ \ \ \ \
D_u = B_s \cap B_t
$$
require the equality of different Cauchy forms presenting the same
invariant amplitude in different hyperlayers
$B_x$ $(x=s,t,u)$.
This will result in appearing of an infinite system of bootstrap
conditions constraining the allowed values of fundamental observables
of a theory (triple coupling constants and mass parameters). Besides,
this system will also completely determine the allowed form of the
4-leg pointlike vertex.
\item
Finally, choose those bootstrap constraints which can be compared with
presently known data and perform the numerical testing.
\end{enumerate}

Below we literally follow this step-by-step instruction and show all
the details of corresponding calculations. This will allow us to omit
these details in subsequent publications devoted to the analysis of
more sophisticated cases.

In this paper we consider a concrete process and employ experimental
data. Thus it is natural to take account of certain well established
phenomenology already on the stage of constructing the amplitude. For
this reason we take the isotopic invariance as an exact symmetry of
strong interaction. Such restrictions are kept automatically when one
uses experimental data to verify theoretical results. On the other
hand, they do not affect the mathematical scheme developed in
\cite{KSAVVV2}
and can easily be relaxed if necessary. Note that we suggest the
absence of massless hadrons with spin
$J \geq 1$
which our technique cannot handle so far. This suggestion is also
supported by experiment.


\section{Structure of the amplitude and resultant vertices}
\label{sec-General}

The amplitude
$M_{a \alpha}^{b \beta}$ of the reaction
$$ 
\pi_{a} (k) + N_{\alpha} (p, \lambda) \to
\pi_{b} (k') + N_{\beta} (p', \lambda')
$$ 
can be presented in the following form:
\begin{equation}
M_{a \alpha}^{b \beta} =
\left\{
\delta_{ba}\delta_{\beta\alpha} M^+ +
i\varepsilon_{bac}(\sigma_{c})_{\beta\alpha} M^{-}
\right\}.
\label{2.2}
\end{equation}
Here
\begin{equation}
M^{\pm} = \overline{u}(p',\lambda') \left\{
A^{\pm}+\left(\frac{\s{k}+\s{k'}}{2}\right) B^{\pm}
\right\} u(p,\lambda)\;\; ,
\label{M-plus-minus}
\end{equation}
$a, b, c = 1,2,3$
and
$\alpha,\beta = 1,2$
stand for the isospin indices,
$\lambda, \lambda'$ ---
for polarizations of the initial and final nucleons, respectively,
$\overline{u}(p',\lambda')$,
$u(p,\lambda)$ ---
for Dirac spinors,
$\sigma_c$ ---
for Pauli matrices:
$$
\left[ {\sigma}_a,\,{\sigma}_b \right]_- =
2\, i\, {\varepsilon}_{abc}\, {\sigma}_c\, ,
$$
and
$\s{p} \equiv p_\mu \gamma^\mu$.
The invariant amplitudes
$A^{\pm}$ and $B^{\pm}$
are considered depending on arbitrary pair of Mandelstam variables
$$
s\equiv (p+k)^2,\ \ \ \ \ \
t\equiv (k-k')^2,\ \ \ \ \ \
u\equiv (p-k')^2.
$$

\begin{figure*}
\includegraphics[height=2.5cm]{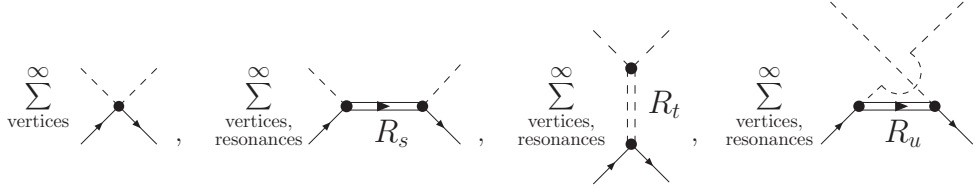}
\caption{\label{1f} Tree level graphs.
$R_s$,
$R_t$
and
$R_u$
stand for all admissible resonances in
$s$-, $t$-, and $u$-channels,
respectively; the formal summation over all possible kinds of vertices
and internal lines is implied.}
\end{figure*}

To compute the tree level expressions for
$A^{\pm}$
and
$B^{\pm}$
one needs to collect contributions from the graphs shown in
Fig.~\ref{1f}.




For this one needs to specify the form of minimal propagators and
resultant triple vertices of three kinds: pion-pion-meson
($\pi \pi M$),
antinucleon-nucleon-meson
(${\overline N}NM$)
and pion-nucleon-baryon
($\pi N B$).
There is no need in explicit parametrization of the resultant
pointlike vertex
$N{\overline N}\pi \pi$
because, as shown in
\cite{KSAVVV2},
its contribution turns out to be entirely fixed by the first
kind bootstrap conditions.

The inner lines of graphs in
Fig.~\ref{1f}
may correspond to mesons (dashed) and baryons (solid).
There are two families of meson resonances which can give a
contribution. The first one contains those with isospin
$I=0$,
even spin
$J=0,2,...$
and positive parity
$P=+1$.
We denote the corresponding fields %
\footnote{We use the Rarita-Schwinger formalism
\cite{Rarita}.}
as
$S_{\mu_1 \ldots \mu_J}$.
The second meson family contains isovector resonances
($I=1$)
with odd spin values
$J=1,3,...$
and negative parity
$P=-1$;
their fields we denote as
$V^a_{\mu_1\ldots\mu_J}$
($a=1,2,3$)
(when forming scalar and vector products we omit the isospin indices
and write isotopic vectors in boldface).

It is convenient to classify possible baryon resonances
according to their
{\em normality}
${\cal N}$:
$$
{\cal N} \equiv (-1)^{(J-1/2)}P.
$$
Here spin
$J=l+1/2$
$(l=0,1,\ldots )$.
Therefore, only four families of baryon resonances contribute to the
amplitude under consideration. We denote them as follows
($\alpha   = 1,2$
and
$a=1,2,3$
stand for the isotopic indices; spinor indices are omitted):
\begin{eqnarray*}
&& (I={1/2},\ \ {\cal N}=+1)\ \ \Longrightarrow\ \ \
R^{\alpha}_{\mu_1\ldots\mu_l};\ \ \ \ \ \ \ \ \ \ \ \nonumber\\&&
(I={3/2},\ \ {\cal N}=+1)\ \ \Longrightarrow\ \ \
\Delta^{a\alpha}_{\mu_1\ldots\mu_l};
\end{eqnarray*}
\begin{eqnarray*}
&&(I={1/2},\ \ {\cal N}=-1)\ \ \Longrightarrow\ \ \
{\widehat{R}}^{\alpha}_{\mu_1\ldots\mu_l};\ \ \ \ \ \ \ \ \ \ \
\nonumber\\&&
(I={3/2},\ \ {\cal N}=-1)\ \ \Longrightarrow\ \ \
{\widehat{\Delta}}^{a\alpha}_{\mu_1\ldots\mu_l}.
\end{eqnarray*}
For example, the famous
$\Delta(1232)$
resonance
($I,J = 3/2$, $P = +1$)
has negative normality; in our notations it belongs to the family
$\widehat{\Delta}$.
Also, it should be kept in mind that the lightest particle with
$l=0$
(spin
$J=1/2$)
in the family
$R$
is just a nucleon.

The resultant vertices are defined and can be properly written down
in momentum space only
\cite{KSAVVV2}.
However, the 3-leg resultant vertices provide an exception; they can
be read from the following Hamiltonian monomials (we use
$\gamma_5 = -i\gamma^0\gamma^1\gamma^2\gamma^3$):
\begin{equation}
H(\pi N R) =
ig_{\sss R}
\overline{N} \bo{\sigma} \gamma_5
R_{\mu_1\ldots\mu_l}
\partial^{\mu_1}\!\!\!\!\ldots\partial^{\mu_l}\bo{\pi}
+ H.c.;
\label{H-piNR}
\end{equation}
\begin{equation}
H(\pi N \widehat{R}) =
g_{\sss \widehat{R}}
\overline{N}\bo{\sigma}
\widehat{R}_{\mu_1\ldots\mu_l}
\partial^{\mu_1}\!\!\!\!\ldots\partial^{\mu_l}\bo{\pi}
+ H.c.;
\label{H-piNRhat}
\end{equation}
\begin{equation}
H(\pi N \Delta) =
ig_{\sss \Delta}
\overline{N} \gamma_5 P_{\sss 3/2}
{\bo{\Delta}}_{\mu_1\ldots\mu_l}
\partial^{\mu_1}\!\!\!\!\ldots\partial^{\mu_l} \bo{\pi}
+ H.c.;
\label{H-piNdelta}
\end{equation}
\begin{equation}
H(\pi N \widehat{\Delta}) =
g_{\sss \widehat{\Delta}}
\overline{N} P_{\sss 3/2}
{\bo{\widehat{\Delta}}}_{\mu_1\ldots\mu_l}
\partial^{\mu_1}\!\!\!\!\ldots\partial^{\mu_l} \bo{\pi}
+ H.c.;
\label{H-piNdeltahat}
\end{equation}
\begin{equation}
H(S\pi\pi) =
\frac{1}{2}\; g_{\sss S\pi\pi}
S_{\mu_1\ldots\mu_J}
(\bo{\pi}\cdot\partial^{\mu_1}\!\!\!\!\ldots\partial^{\mu_J}\bo{\pi})
\; ;
\label{H-Spipi}
\end{equation}

\begin{eqnarray}
\label{H-SNN}
H(SNN)=&&
\left[
g^{(1)}_{\sss NNS}
\overline{N}\partial_{\mu_1}\!\!\ldots\partial_{\mu_J} N
\right.
\nonumber\\
&&
+
\left. ig^{(2)}_{\sss NNS} J
\partial_{\mu_1}\!\!\ldots\partial_{\mu_{J-1}}
\overline{N}\gamma_{\mu_J} N
\right]
S^{\mu_1\ldots\mu_J} \nonumber\\
\end{eqnarray}

\begin{equation}
H(V\pi\pi) =
\frac{1}{2}\; g_{\sss V\pi\pi}
\bo{V}_{\mu_1 \ldots \mu_J}
(\bo{\pi}\times\partial^{\mu_1}\!\!\!\!\ldots\partial^{\mu_J}\bo{\pi})
\; ;
\label{H-Vpipi}
\end{equation}
\begin{eqnarray}
\label{H-VNN}
H(VNN) = &&
\left[
i g^{(1)}_{\sss NNV}
\overline{N} \bo{\sigma}
\partial_{\mu_1}\!\!\ldots\partial_{\mu_J} N
\right.
\nonumber\\
&&
+
\left. g^{(2)}_{\sss NNV} J \overline{N} \gamma_{\mu_J}
\bo{\sigma} \partial_{\mu_1}\!\!\ldots\partial_{\mu_{J-1}} N
\right]
\bo{V}^{\mu_1\ldots\mu_J}\; . \nonumber\\
\end{eqnarray}
In
Eqs.~(\ref{H-piNdelta}, \ref{H-piNdeltahat})
$P_{\sss 3/2}$
denotes the isospin-3/2 projecting operator:
\begin{eqnarray}
P_{\sss 3/2} \equiv
\left(P_{\sss 3/2}\right)_{a \alpha b \beta} = &&
\frac{2}{3} \left\{
\delta_{\alpha\beta}\delta_{ab} -
\frac{i}{2}\; \varepsilon_{abc}
\left( \sigma_c \right)_{\alpha\beta}
\right\},\ \ \ \ \ \nonumber\\ &&
(a, b = 1,2,3;\; \alpha,\beta=1,2).
\end{eqnarray}

One can easily check that in momentum space these monomials provide
the full set of 3-leg minimal vertices under the condition that the
independent variables are chosen as
$p_n^2$
where
$p_n$
($n=1,2,3$)
stands for the 4-momentum of
$n$th
leg.

The 0th level coupling constants that appear in equations
(\ref{H-piNR}) -- (\ref{H-VNN})
are real.
According to the results of
\cite{KSAVVV2}
listed in Section
\ref{sec-Preliminaries}
these couplings present the fundamental physical observables.

The general form of the minimal propagator of a particle with  mass
parameter
$M$
and spin number
$l$
(this corresponds to spin
$J=l$
for boson and
$J=l+1/2$
for fermion) looks as follows:
\begin{equation}
P^{\mu_1 \ldots \mu_l}_{\nu_1 \ldots \nu_l}(q;l)=
\frac{i}{(2\pi)^4}\;
\frac{\Pi^{\mu_1 \ldots \mu_l}_{\nu_1 \ldots
                 \nu_l}(q;l)}{q^2 - M^2 + i\epsilon}\;\; .
\label{general-propagator}
\end{equation}
Here
$
\Pi^{\mu_1 \ldots \mu_l}_{\nu_1 \ldots \nu_l}(q;l)
$
is the relevant spin sum constructed from the Rarita-Schwinger wave
functions
$
{\cal E}^{\mu_1 \ldots \mu_l}(i,p)
$
and defined in
(\ref{spin-sum-boson})
for bosons and in
(\ref{spin-sum-fermion})
for baryons.
The eventual spinor indices and isotopic factors like
$\delta_{ab}$, $\delta_{\alpha \beta}$
and
$P_{3/2}^{a\alpha b\beta}$
are omitted. The main properties of such spin sums are summarized in
the Appendix~\ref{sec-contproj}.

Now we have in hand all the ingredients needed to calculate those
elements of tree level graphs which are used for constructing the
Cauchy forms. In the next Section we explain certain specific details
of the computational procedure.

\section{Resonance exchange graph: example of computation}
\label{sec-example}

To construct the Cauchy forms for the scalar amplitudes
$A^{\pm}$ and $B^{\pm}$
in
(\ref{M-plus-minus}),
one needs to know the residues at the relevant resonance poles. Below
we demonstrate how the contracted projector formalism (briefly
reviewed in Appendix~\ref{sec-contproj})
allows one to compute the contributions to these residues that follow
from graphs with arbitrary spin resonance%
\footnote{When speaking about internal lines we often use the term
``resonance''
for both stable and unstable particles.}
exchanges.

\begin{figure}
\includegraphics[height=2.1cm]{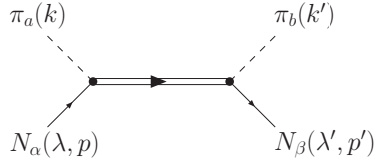}
\caption{\label{fr1f} Typical graph with a fermion resonance exchange.
Here
$a,b,\alpha,\beta$
stand for isotopic indices and
$\lambda, \lambda'$ ---
for nucleon polarizations.}
\end{figure}

As an example, consider the graph
(Fig.~\ref{fr1f})
corresponding to the
$s$-channel
exchange by a resonance with spin
$J=l+1/2$,
isospin
$I=1/2$
and negative normality
${\cal N}=-1$.


The left and right resultant vertices are easily read from
(\ref{H-piNRhat}).
They are, respectively,
\begin{equation}
-i\, g^{\ast}_{\sss \widehat{R}} (-i)^l k^{\nu_1} \ldots k^{\nu_l}
(\sigma_a)_{\gamma\alpha}
\label{sample-vertex-left}
\end{equation}
and
\begin{equation}
-i\, g^{}_{\sss \widehat{R}} (i)^l {k'}_{\mu_1} \ldots {k'}_{\mu_l}
(\sigma_b)_{\beta\gamma}
\label{sample-vertex-right}
\end{equation}
($\gamma = 1,2$
is the isotopic index of the resonance). The corresponding minimal
propagator is given by the expression
(\ref{general-propagator})
with
$l=J-1/2$
($\rho$, $\tau$
stand for spinor indices
and
$M$ --
for the resonance mass parameter).

With the help of
(\ref{general-propagator}),
(\ref{sample-vertex-left}), (\ref{sample-vertex-right})
and
(\ref{cpr5})
one can write down the contribution of the resultant graph shown on
Fig.~\ref{fr1f}
to the amplitude of elastic pion-nucleon scattering as follows:
\begin{eqnarray}
\label{sample-resultant-contribution}
&& G_{b \beta\, a\alpha}(p,k,\lambda;\ p',k',\lambda ') =
\nonumber\\ &&
({\sigma_b\, \sigma_a})^{}_{\beta \alpha}\;
g^{\ast}_{\sss \widehat{R}}\;g^{}_{\sss \widehat{R}}\;
\overline{u} (p',\lambda')\;
{\cal P}^{(l+\frac{1}{2})}(k',k,k+p)\;
u(p,\lambda)\; ,
\nonumber\\
\end{eqnarray}
where
$u(p,\lambda)$
and
$\overline{u} (p',\lambda')$
stand for the nucleon wave functions and
$
{\cal P}^{(l+\frac{1}{2})}(k',k,k+p)
$
--- for contracted projector. Finally, using  the explicit
form
(\ref{contracted-projector-fermion-explicit})
of the contracted projector, one obtains the following expression
for the contribution of the graph under consideration:
\begin{widetext}
\begin{equation}
G_{b \beta\, a\alpha}(\ldots) =
- \frac{G(\pi N \widehat{R})}{s- M_{\sss \widehat{R}}^2}\;
({\sigma_b\, \sigma_a})^{}_{\beta \alpha}\;
\overline{u} (p',\lambda')
\left[
F^l_{\sss A} (M,t) +
\frac{\left( \s{k}+ \s{k}' \right)}{2} F^l_{\sss B} (M, t)
\right]
u(p,\lambda).
\label{fr8}
\end{equation}
\end{widetext}
Here
\begin{equation}
\label{G-pi-N-R-hat}
G(\pi N \widehat{R}) \equiv |g^{}_{\sss \widehat{R}}|^2\;
{\Phi}^l\, \frac{l!}{(2l+1)!!}\; ,
\end{equation}
\begin{eqnarray}
\label{Kallen-function-Phi}
&& \Phi (M,m,\mu) \equiv \nonumber\\ &&
\frac{1}{4M^2} \left( M^4+m^4+\mu^4-2M^2m^2-2M^2\mu^2-2m^2\mu^2
\right)\; , \nonumber\\
\end{eqnarray}
and
$m$, $\mu$
stand for the nucleon and pion mass, respectively. Two auxiliary
functions
$F^l_{\sss A} (M,t)$
and
$F^l_{\sss B} (M,t)$
are defined as follows:
\begin{eqnarray}
\label{function-FAl}
F_{\sss A}^{l}(M,t) \equiv &&
(M+m) P'_{l+1}\left(1+\frac{t}{2\Phi}\right)  \nonumber\\ && +
(M-m) \frac{(M+m)^2-\mu^2}{(M-m)^2-\mu^2}
P'_l\left(1+\frac{t}{2\Phi}\right)\; ,
\nonumber\\
\end{eqnarray}
\begin{eqnarray}
\label{function-FBl}
F_{\sss B}^{l}(M,t) \equiv &&  P'_{l+1}\left(1+\frac{t}{2\Phi}\right)
\nonumber\\ &&
-\frac{(M+m)^2-\mu^2}{(M-m)^2-\mu^2}
P'_l\left(1+\frac{t}{2\Phi}\right)\; .
\end{eqnarray}

Comparing now
(\ref{fr8})
with
(\ref{2.2})
and using the well known relation for Pauli matrices
$$
({\sigma_b\, \sigma_a})^{}_{\beta \alpha} =
\delta_{ba} \delta_{\beta \alpha} +
i\, \varepsilon_{bac} (\sigma_c)_{\beta \alpha}\; ,
$$
we conclude that the graph on
Fig.~\ref{fr1f}
gives the following contributions to the singular (or, the same,
{\em principal})
parts of invariant amplitudes:
\begin{eqnarray*}
&& {\rm to}\; A^{+} \;: \quad
-\frac{G(\pi N \widehat{R})}{s-M_{\sss \widehat{R}}^2}\; F^l_{\sss A}(M,t)\;
,
\nonumber\\ &&
{\rm to}\; A^{-} \;: \quad
-\frac{G(\pi N \widehat{R})}{s-M_{\sss \widehat{R}}^2}\; F^l_{\sss A} (M,t)\; ,
\end{eqnarray*}
\begin{eqnarray*}
&&
{\rm to}\; B^{+} \;: \quad
-\frac{G(\pi N \widehat{R})}{s-M_{\sss \widehat{R}}^2}\; F^l_{\sss B} (M,t)\; ,\ \
\nonumber\\ &&
{\rm to}\; B^{-} \;: \quad
-\frac{G(\pi N \widehat{R})}{s-M_{\sss \widehat{R}}^2}\; F^l_{\sss B} (M,t)\; .
\end{eqnarray*}

In the same way, using the relations
(\ref{contracted-projector-boson-explicit})
and
(\ref{contracted-projector-fermion-explicit})
one can derive expressions for all the other resultant graphs which
correspond to a resonance exchange in one of the channels
(see Fig.~\ref{1f}).
The results are listed in
Appendix~\ref{app-graphs}.
This fixes the principal parts of tree level invariant amplitudes.


\section{Constructing the Cauchy forms}
\label{sec-Cauchy}

\begin{figure}
\includegraphics[height=7.0cm]{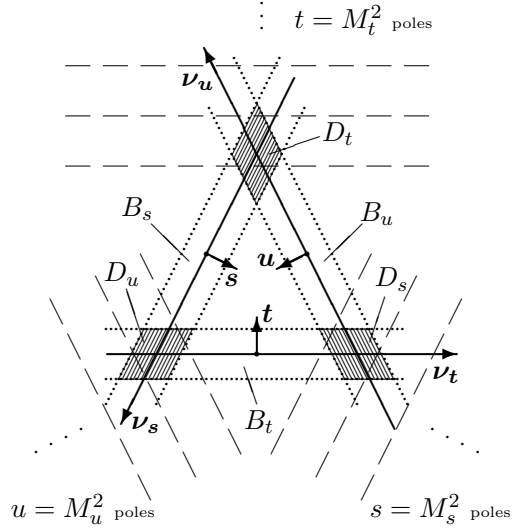}
\caption{\label{5f} Mandelstam plane: three different Cauchy series
uniformly converge in three different hyperlayers
$B_s$, $B_t$ and $B_u$
(the projections are bounded by dotted lines). The intersection
domains
$D_s$, $D_t$, $D_u$
are hatched. The dashed lines show the pole positions in the relevant
channels.}
\end{figure}

In this Section we construct the Cauchy forms for tree level
amplitudes
$A^{\pm}$
and
$B^{\pm}$
in three hyperlayers
$B_s$, $B_t$ and $B_u$
(their projections on the Mandelstam plane are shown on
Fig.~\ref{5f}).

To construct the Cauchy form presenting a given tree level amplitude
in a certain hyperlayer, one needs to know the degree of relevant
bounding polynomial, the configuration of poles, and the explicit
expressions for corresponding residues. The location of poles is fixed
(or, better, parameterized) with the help of the mass parameters
$M_i$.
The corresponding residues are listed in the
Appendix~\ref{app-graphs}.

The bounding polynomial degrees are chosen in accordance with known
values of corresponding Regge intercepts
(see Sec.~\ref{sec-Preliminaries}).
In all the cases we have already examined
($\pi\pi$, $\pi K$, $\pi N$,
and
$KN$
elastic scattering processes) it happens impossible to obtain
reasonable (at least, roughly corresponding to known data) bootstrap
conditions until this requirement is fulfilled.
In the reaction under consideration the intercepts are
(see e.g.
\cite{Collins1977}):
\begin{equation}
\alpha^\mathrm{M}_{\sss 0} = 1\, ;\ \ \ \ \
\alpha^\mathrm{M}_{\sss 1} \sim 0.5\, ;\ \ \ \ \
\alpha^\mathrm{B}_{\sss 1/2} \sim 0\, ;\ \ \ \ \
\alpha^\mathrm{B}_{\sss 3/2} < 0\, ;
\label{intercepts}
\end{equation}
(here the upper indices
$\mathrm{M}$
and
$\mathrm{B}$
correspond to meson and baryon trajectory, respectively, while the
lower ones refer to the isospin value). Using the numerical values
(\ref{intercepts})
we conclude that:

$\bullet$
in $B_s$:
\begin{eqnarray}
\label{degree1-in-Bs}
&& \left( A^+ + 2 A^- \right)
\Bigr|_{|\nu_s| \rightarrow \infty}\;
{\sim}\; O\left( \frac{1}{|\nu_s|} \right)\, ,
\nonumber\\
&&\left( B^+ + 2 B^- \right)
\Bigr|_{|\nu_s| \rightarrow \infty}\;
{\sim}\; O\left( \frac{1}{|\nu_s|} \right);
\end{eqnarray}
\begin{eqnarray}
&& \left( A^+ - A^- \right)
\Bigr|_{|\nu_s| \rightarrow \infty}\;
{\sim}\; o\left( \frac{1}{|\nu_s|} \right)\,,
\nonumber\\&&
\left( B^+ - B^- \right)
\Bigr|_{|\nu_s| \rightarrow \infty}\;
{\sim}\; o\left( \frac{1}{|\nu_s|} \right);
\label{degree2-in-Bs}
\end{eqnarray}

$\bullet$
in $B_t$:
\begin{eqnarray}
&&\left( A^+ \right)
\Bigr|_{|\nu_t| \rightarrow \infty}\;
{\sim}\; o\left( |\nu_t|^2 \right)\, ,
\nonumber\\&&
\left( B^+ \right)
\Bigr|_{|\nu_s| \rightarrow \infty}\;
{\sim}\; O(1)\, ;
\label{degree1-in-Bt}
\end{eqnarray}
\begin{eqnarray}
&&\left(A^- \right)
\Bigr|_{|\nu_t| \rightarrow \infty}\;
{\sim}\; o\left( |\nu_t| \right)\, ,
\nonumber\\&&
\left(B^- \right)
\Bigr|_{|\nu_s| \rightarrow \infty}\;
{\sim}\; o(1)\, ;
\label{degree2-in-Bt}
\end{eqnarray}

$\bullet$
in $B_u$:
\begin{eqnarray}
&& \left( A^+ - 2 A^- \right)
\Bigr|_{|\nu_s| \rightarrow \infty}\;
{\sim}\; o(1)\, ,
\nonumber\\ &&
\left( B^+ - 2 B^- \right)
\Bigr|_{|\nu_s| \rightarrow \infty}\;
{\sim}\; o(1)\, ;
\label{degree1-in-Bu}
\end{eqnarray}
\begin{eqnarray}
&&
\left( A^+ - A^- \right)
\Bigr|_{|\nu_s| \rightarrow \infty}\;
{\sim}\; o(1)\, ,
\nonumber\\ &&
\left( B^+ - B^- \right)
\Bigr|_{|\nu_s| \rightarrow \infty}\;
{\sim}\; o(1)\, .
\label{degree2-in-Bu}
\end{eqnarray}

Thus in
$B_s$
and
$B_u$
the invariant amplitudes
$A^{\pm}$
and
$B^{\pm}$
possess decreasing  asymptotics. Therefore
(see \cite{KSAVVV2})
one does not need to take account of any correcting polynomials and
subtraction terms in the Cauchy forms valid in these layers. The same
is true with respect to
$B^-$
in
$B_t$.
Next, since
$A^-$ and $B^+$
are odd functions of
$\nu_t$
(this is just a consequence of Bose symmetry),
the zero degree is ruled out here and the correcting polynomials can
also be dropped as well as the subtraction terms. At last, because
$A^+$
is even in
$B_t$,
the equation
(\ref{degree1-in-Bt})
tells us that the degree of corresponding bounding polynomial is zero.

We conclude that the only Cauchy form which requires taking account of
(0th order in
$\nu_x$)
correcting polynomials and the corresponding substraction term is that
representing the invariant amplitude
$A^+(t,\nu_t)$
in the hyperlayer
$B_t$.
In all other cases neither background terms nor correcting polynomials
are needed; the corresponding Cauchy series are just the properly
ordered sums of pole terms stemming from the relevant resultant
graphs.

Now we can construct the Cauchy forms which provide the uniformly
converging series for invariant amplitudes
$A^{\pm}$ and $B^{\pm}$
in three hyperlayers
$B_s$, $B_t$ and $B_u$.
However, the explicit expressions are too bulky. To make them
readable we need to introduce more compact notations.

Henceforth
$X^{\pm}$
stands for
$A^{\pm}$
or
$B^{\pm}$,
$M$
denotes the relevant resonance (baryon or meson) mass parameter and,
as usual,
$m$ ($\mu$)
is the nucleon (pion) mass. Further, introducing the abbreviation
\begin{equation}
 C_{\sss I}^{\pm}:
\left\{
C^{+}_{\sss 1/2} = 1;\;
C^{-}_{\sss 1/2} = 1;\;
C^{+}_{\sss 3/2} = 2/3;\;
C^{-}_{\sss 3/2} =-1/3
\right\}\, ,
\label{C-I}
\end{equation}
we define for baryons
\begin{equation}
Y^{\pm}_{\sss X} (M, \chi) \equiv
\sum_{I= 1/2, 3/2
\atop
J = 1/2, 3/2, \ldots} C_{\sss I}^{\pm}
G(\pi N {\cal R}) F_{\sss X}^{\: l}(-{\cal N}M, \chi)\, ,
\label{YXpm}
\end{equation}
where
$\chi$
stands for arbitrary kinematical variable and
${\cal N}$ --
for normality, the summation being implied over all baryon resonances
(${\cal R} = R, \widehat{R}, \Delta, \widehat{\Delta}$)
with the same mass
$M$.
Similarly, for mesons:
\begin{equation}
W_{\sss X}^{\pm} (M,\chi) \equiv
\sum_{\stackrel{\scriptstyle I= 0,1}{\scriptstyle J= 0,1,\ldots}}^{}
\frac{1}{2}\left[ 1\pm (-1)^I \right]
W_{\sss X}(I,J,\chi)\; ,
\label{WXpm}
\end{equation}
where
\begin{eqnarray}
\label{W-A}
W_{\sss A}(I,J,\chi)
\equiv
&& \frac{1}{2}\left[ (-1)^I + (-1)^J \right]
\nonumber\\ &&
\times \left\{
G^I_1\, P_{\sss J} (\chi) 
-\frac{4m}{4m^2 - M^2}\; G^I_2\; P'_{\sss J-1} (\chi)
\right\},
\nonumber\\
\end{eqnarray}
\begin{equation}
W_{\sss B}(I,J,\chi)
\equiv
\frac{1}{2} \left[(-1)^I + (-1)^J \right] \frac{1}{F}\;
G^I_2\; P'_J (\chi) .
\label{W-B}
\end{equation}
In the
Eq.~(\ref{WXpm})
the summation%
\footnote{
Both sums in
(\ref{YXpm})
and
(\ref{WXpm})
are {\em finite}
because, as mentioned in
Sec.~\ref{sec-General},
we imply that the number of resonances with the same value of mass
parameter is finite. To put it another way, we imply existence of the
leading Regge trajectory (in the plane
$(J,M)$)
which, however, is not necessarily linear.
}
is implied over all non-strange meson resonances with the same mass
$M$
and natural parity
$P = (-1)^J$.
Finally, introducing the sign regulator
\begin{equation}
\eta_{\sss X} =
\left\{
\begin{array}{ll}
+1, & X=A\\
-1, & X=B
\end{array}
\right.
\label{sign-regulator}
\end{equation}
and abbreviations
\begin{equation}
\Sigma \equiv [M^2-2(m^2 +\mu^2)]\, ,\ \ \ \ \ \ \ \
\theta \equiv (M^2-m^2-\mu^2)\, ,
\label{Sigma-theta}
\end{equation}
we can write down the compact expressions for desired Cauchy forms.

\noindent
$\bullet$
$B_s\{\nu_s \in {\mathbb{ C}};\, s \in {\mathbb{R}},\, s \sim 0 \}$.

Here the relevant poles are those in
$t$
and
$u$.
The asymptotic behavior of every invariant amplitude
$A^{\pm}(s,\nu_s)$
and
$B^{\pm}(s,\nu_s)$
in
$\nu_s$
corresponds to the negative degree of bounding polynomial. Thus we see
that each one of these amplitudes can be presented as follows
$(X=A,B)$:
\begin{align}
X^{\pm}\Big|_{B_s} & =
(\pm \eta_{\sss X}) \sum_{\rm baryons}^{}Y^{\pm}_{\sss X}
\Bigl( M,\, -(\Sigma + s)\Bigr)
\frac{-\, 2}{\nu_s - (s+2\theta)} \notag \\
& + \sum_{\rm mesons}^{}W^{\pm}_{\sss X}
\left( M,\, \frac{\Sigma + 2s}{4F} \right)
\frac{2}{\nu_s + (s+2\theta)}\ \ .
\label{CauchyBs}
\end{align}

\noindent
$\bullet$
$B_t\{\nu_t \in {\mathbb{C}}; \, t \in {\mathbb{R}}, \, t \sim 0 \}$.

As mentioned above, in this hyperlayer the amplitude
$A^+$
requires accounting for the 0th degree correcting polynomials and
subtraction term. With the latter term denoted as
$\alpha(t)$
the correct Cauchy form reads:
\begin{eqnarray}
\label{ACauchyBt}
A^{+}\Big|_{B_t} =
\alpha(t) 
-&&\sum_{\rm baryons}^{}Y^{+}_{\sss A}(M,t)
\left[
\frac{2}{\nu_t-(t+2\theta)} \right.
\nonumber\\ &&
\left.
-\frac{2}{\nu_t+(t+2\theta)} +
\frac{4}{t+2\theta}
\right]\; .
\end{eqnarray}
At the same time, the amplitudes
$A^-$
and
$B^{\pm}$
do not require accounting for correcting polynomials. Hence the
relevant Cauchy forms read:
\begin{eqnarray}
\label{CauchyBt}
&& A^{-}\Big|_{B_t} =
\nonumber\\&&
- \sum_{\rm baryons}^{}Y^{-}_{\sss A} (M,t)
\left[
\frac{2}{\nu_t-(t+2\theta)} - 
\frac{-2}{\nu_t+(t+2\theta)}
\right]\; ,
\nonumber\\ &&
B^{\pm}\Big|_{B_t}   =
\nonumber\\&&
- \sum_{\rm baryons}^{}Y^{\pm}_{\sss B} (M,t)
\left[
\frac{2}{\nu_t-(t+2\theta)} \mp 
\frac{-2}{\nu_t+(t+2\theta)}
\right]\; .
\end{eqnarray}

\noindent
$\bullet$
$B_u\{\nu_u \in {  \mathbb{C}}; \, u \in {  \mathbb{R}}, \, u \sim 0 \}$.

In this hyperlayer the situation is analogous to that in
$B_s$.
Thus we have
$(X=A,B)$:
\begin{align}
X^{\pm}\Big|_{B_u} = & -\sum_{\rm baryons}^{}Y^{\pm}_{\sss X}
\Bigl( M , -(\Sigma + u) \Bigr)
\frac{-2}{\nu_u + (u+2\theta)} \notag \\
& - \sum_{\rm mesons}^{}W^{\pm}_{\sss X}
\left( M ,- \frac{\Sigma+2u}{4F} \right)
\frac{2}{\nu_u - (u+2\theta)}\ \ .
\label{CauchyBu}
\end{align}

We would like to stress that all the sums over resonance contributions
should be taken in order of increasing mass --- otherwise the
convergence of the Cauchy series cannot be guaranteed
(see, e.g.,
\cite{KSAVVV2}). The formal
separation of these sums into meson and baryon parts is done just to
show the explicit form of both kinds of contributions.

Before proceeding further it is useful to summarize briefly what has
been done up to this moment.

First, we performed the classification of all the minimal triple
vertices that describe the interaction of pions and nucleons with
meson and baryon resonances of arbitrary high spin
$J$
and isospin
$I \leq 3/2$.%

Second, we have calculated the explicit form of the residues at poles
stemming from graphs that correspond to resonance exchanges in one of
three channels of the considered process. This allowed us to
separate the full collection of contributions from the tree level
graphs
(Fig.~\ref{1f})
into two
{\em formal}
infinite sums, the first one being solely constructed from the pole
terms while the second is a (formal) power series in arbitrary pair of
independent kinematical variables
$(x, \nu_x)$.

Third, following the procedure proposed in
\cite{AVVV2, KSAVVV2}
(and suggestions listed in
Sec.~\ref{sec-Preliminaries}),
we constructed the uniformly converging Cauchy series
(\ref{CauchyBs}) -- (\ref{CauchyBu})
which provide the correct forms of invariant amplitudes in three
hyperlayers
$B_x$.
Let us stress that these series are constructed from the well-defined
expressions, the only item still unspecified being the subtraction
term
$\alpha (t)$
that appears in
(\ref{ACauchyBt}).

The important feature of the Cauchy forms
(\ref{CauchyBs}) -- (\ref{CauchyBu})
is that, as a rule, neither poles in
$x$
nor smooth (`background') terms depending on both variables
$(x, \nu_x)$
appear explicitly in a form valid in
$B_x$.
The only exception is the Cauchy form
(\ref{ACauchyBt})
for
$A^+$
in
$B_t$.
It contains the background term
$\alpha (t)$
depending on
$t$.
This means that there must exist a mutual cancellation between the
direct channel background terms and the cross channel poles, this
cancellation being complete in all the hyperlayers except
$B_t$.
In this latter case the remnant of cross channel poles and background
contributions survives in the amplitude
$A^+(t, \nu_t)$.
It manifests itself in a form of (still unspecified) subtraction term
$\alpha (t)$
and an infinite number of well-defined smooth terms (the correcting
polynomials) that appear in each item of the sum over pole
contributions.

Such a cancellation might seem a miracle if ever possible since it
requires extremely fine tuning of the structure of a set of resultant
parameters. Fortunately, there exists an example which allows one to
trace the mechanism of this phenomenon --- the famous string amplitude
based on Eyler's B-function. This example has been analyzed in
\cite{POMI, AVVV1}.
It was shown that the corresponding bootstrap conditions present
nothing but an infinite set of identities for Pochhammer symbols which
easily undergo numerical verification.

For this reason it is interesting to construct the explicit form of
bootstrap conditions for
$\pi N$
scattering amplitude and compare them with known data. As mentioned
above, this may provide a test of consistency of the set of
requirements listed in
Sec.~\ref{sec-Preliminaries}.

\section{Bootstrap conditions}
\label{sec-bootstrap}

According to the analysis presented in
\cite{AVVV2}-\cite{POMI},
the full system of bootstrap restrictions is a system of necessary
conditions limiting the values of resultant parameters of various
levels. This system ensures self-consistency of the effective theory
in
$S$-matrix
sector. It arises as a direct consequence of the summability
principle.

Below we consider only a small (though infinite!) part of this system.
Namely, we derive the
{\em tree level bootstrap restrictions}
for the resultant parameters of
$\pi N$
scattering (masses and triple coupling constants that appear in
(\ref{H-piNR}) -- (\ref{H-VNN})).
As argued in
\cite{KSAVVV2},
the bootstrap conditions possess the property of renormalization
invariance: irrelatively to their loop level, they restrict the
possible values of physical observables in a given effective theory.
It is for this reason that already the tree-level bootstrap conditions
can be verified by the direct comparison with experimental data.

First of all let us derive the bootstrap condition which allows one to
express the unknown function
$\alpha (t)$
(see
Eq.~(\ref{ACauchyBt}))
in terms of triple couplings and masses. It follows from the existence
of
{\em two}
Cauchy forms (namely,
(\ref{ACauchyBt})
and
(\ref{CauchyBu}))
which present the same function
$A^+$,
both forms being valid in the domain
$D_s = B_u \cap B_t$.
Hence in this domain they must coincide identically. This gives:
\begin{widetext}
\begin{eqnarray}
\label{alpha-t}
\alpha (t) =   && \sum_{\rm baryons}^{}
\left\{
\frac{Y^+_A(M,-(u+\Sigma)) - Y^+_A(M, t)}{\Sigma + t + u}
+ Y^{+}_A (M, t)
\left[\frac{1}{u-M^2} + \frac{4}{t + 2\theta} \right]
\right\}       \nonumber\\&&
- \sum_{\rm mesons}^{}W^{+}_A
\left( M , - \frac{(2u + \Sigma)}{4F} \right)
\frac{1}{t-M^2}\, \equiv \Psi_s (A^+)  \;  ;\ \ \ \ \ \ (t,u) \in D_s\; .
\end{eqnarray}
\end{widetext}
Here we have used
(\ref{x_nu})
to express
$\nu_t$
and
$\nu_u$
in terms of
$(t,u)$.
In
\cite{KSAVVV2}
the relations of the type
(\ref{alpha-t})
have been called as the bootstrap conditions of the first kind.

As we have already mentioned in
Sec.~\ref{sec-Cauchy},
$\alpha (t)$
(as well as the correcting polynomials) results from the
contributions of contact (pointlike) graphs and from the graphs with
$t$-channel
resonance exchanges. Nevertheless, the right side of
(\ref{alpha-t})
only depends on the tree level resultant coupling constants at
triple vertices.
Thus the relation
(\ref{alpha-t})
gives an illustration to the general statement made in
\cite{KSAVVV2}:
there is no need ton formulate the independent renormalization
prescriptions for 4-leg amplitudes as long as the true (experimental)
asymptotic behavior is taken into account in our scheme.

The formula
(\ref{alpha-t})
is only valid in
$D_s$;
outside this domain it is meaningless because at least one of two
series
(\ref{ACauchyBt}),
(\ref{CauchyBu})
may diverge. For this reason the pole terms which appear in the
right side, in fact, do not correspond to singularities --- the
function
$\alpha (t)$
is smooth in
$D_s$.
Moreover, since it only may depend on
$t$,
the expression
(\ref{alpha-t})
defines this function everywhere in the hyperlayer
$B_t$
{\em under the condition that the parameters fulfil certain
self-consistency restrictions which provide a guarantee of
independence of the right side on the variable}
$u$.
In the case under consideration these restrictions may be written
as follows%
\footnote{
Here the reference point
$(t,u)=(0,0)$
is chosen just for convenience; in principle, every point
$(t,u) \in D_s$
would be equally acceptable.
}:
\begin{equation}
{\partial}^{\, m+1}_u\, {\partial}^{\, n}_t \Psi_s (A^+)
\Bigr|_{t,u=0} = 0\,\, ,\ \ \ \ \ \
(m,n=0,1,\ldots).
\label{alpha-derivatives}
\end{equation}

The infinite system of equations
(\ref{alpha-derivatives})
only contains the numerical parameters%
\footnote{In
\cite{KSAVVV2}
the systems of this type are called as the second kind bootstrap
conditions.}
--- the resultant triple coupling constants and masses. It provides an
example of sum rules that connect among themselves the parameters of
fermion and boson spectra.

Clearly, the system
(\ref{alpha-derivatives})
presents only one of necessary self-consistency conditions. Indeed,
there are three domains where two of three hyperlayers
($B_s$, $B_t$ and $B_u$)
intersect:
$$
D_s = B_t \cap B_u;\ \ \ \ \ \ \ \
D_t = B_u \cap B_s;\ \ \ \ \ \ \ \
D_u = B_s \cap B_t.
$$
Therefore, we have three systems of such functional self-consistency
conditions, namely:
\begin{equation}
A^{\pm}\Bigr|_{\sss B_t} = A^{\pm}\Bigr|_{\sss B_u},\ \ \ \ \ \ \
B^{\pm}\Bigr|_{\sss B_t} = B^{\pm}\Bigr|_{\sss B_u},\ \ \ \ \ \ \
(t,u) \in D_s;
\label{corner-Ds}
\end{equation}
\begin{equation}
A^{\pm}\Bigr|_{\sss B_u} = A^{\pm}\Bigr|_{\sss B_s},\ \ \ \ \ \ \
B^{\pm}\Bigr|_{\sss B_u} = B^{\pm}\Bigr|_{\sss B_s},\ \ \ \ \ \ \
(u,s) \in D_t;
\label{corner-Dt}
\end{equation}
\begin{equation}
A^{\pm}\Bigr|_{\sss B_s} = A^{\pm}\Bigr|_{\sss B_t},\ \ \ \ \ \ \
B^{\pm}\Bigr|_{\sss B_s} = B^{\pm}\Bigr|_{\sss B_t},\ \ \ \ \ \ \
(s,t) \in D_u.
\label{corner-Du}
\end{equation}
Obviously in the case of
$\pi N$
elastic scattering the systems
(\ref{corner-Du})
and
(\ref{corner-Ds})
are completely equivalent. For this reason it is quite sufficient to
consider only two systems:
(\ref{corner-Ds})
and
(\ref{corner-Dt}).
It is convenient to present them in terms of two groups of generating
functions.

The functions from the first group generate the self-consistency
(bootstrap) conditions
(\ref{corner-Ds}).
We define them as follows%
\footnote{Except
$\Psi_s (A^+)$,
all these functions are just the differences of two relevant Cauchy
forms.}:
\begin{widetext}
\begin{eqnarray}
\label{psi-Aplus-Ds}
\Psi_s (A^+) \equiv
&& \sum_{\rm baryons}^{}
\left\{
\frac{Y_A^+ (M , -(u+\Sigma)) - Y_A^+ (M , t)}{\Sigma + t + u}
+ Y_A^+ (M , t)
\left[ \frac{1}{u-M^2} + \frac{4}{t+2\theta} \right]
\right\}       \nonumber\\ &&
-  \sum_{\rm mesons}^{}
\frac{W_A^+ \left( M , -(2u+\Sigma)/4F \right)}{t-M^2}\, ;
\\
\label{psi-Aminus-Ds}
\Psi_s (A^-) \equiv  &&
\sum_{\rm baryons}^{}
\left\{
\frac{Y_A^- (M , -(u+\Sigma)) - Y_A^- (M , t)}{\Sigma + t + u}
-\frac{Y_A^- (M , t)}{u-M^2}
\right\}
\nonumber\\&&
-\sum_{\rm mesons}^{}
\frac{W_A^- \left( M , -(2u+\Sigma)/4F \right)}{t-M^2}\ ; \\
\label{psi-Bplus-Ds}
\Psi_s (B^+) \equiv  &&
\sum_{\rm baryons}^{}
\left\{
\frac{Y_B^+ (M , -(u+\Sigma)) - Y_B^+ (M , t)}{\Sigma + t + u}
-\frac{Y_B^+ (M , t)}{u-M^2}
\right\}  \nonumber\\ &&
-\sum_{\rm mesons}^{}
\frac{W_B^+ \left( M,-(2u+\Sigma)/4F \right)}{t-M^2}\ ; \\
\label{psi-Bminus-Ds}
\Psi_s (B^-) \equiv  &&
\sum_{\rm baryons}^{}
\left\{
\frac{Y_B^- (M,-(u+\Sigma)) - Y_B^- (M,t)}{\Sigma + t + u}
+\frac{Y_B^- (M,t)}{u-M^2}
\right\}  \nonumber\\&&
-\sum_{\rm mesons}^{}
\frac{W_B^- \left( M,-(2u+\Sigma)/4F \right)}{t-M^2}\ .
\end{eqnarray}
\end{widetext}

The corresponding bootstrap conditions read:
\begin{eqnarray}
\label{bootstrap-Aplus-Ds}
&& {\partial}^m_t\, {\partial}^{n+1}_u\,
\Psi_s (A^+){\Bigr|}_{t,u=0}\, = 0\, ,\ \ \
(m,n=0,1,\ldots); \ \ \ \
\\
&& {\partial}^m_t\, {\partial}^n_u\,
\Psi_s (X_s){\Bigr|}_{t=u=0}\, = 0\, ,
\nonumber\\ &&
(m,n=0,1,\ldots)\, .\ \ \ \
(X=A^-, B^+, B^-).
\label{bootstrap-Aminus-Bplus-Bminus-Ds}
\end{eqnarray}

Similarly, the second group of generating functions is defined as:
\begin{widetext}
\begin{eqnarray}
\label{psi-Aplus-Dt}
\Psi_t (A^{+}) \equiv  &&
\sum_{\rm baryons}^{}
\left[
\frac{Y_A^+ (M,-(u+\Sigma))}{s-M^2}
-\frac{Y_A^+ (M,-(s+\Sigma))}{u-M^2}
\right]  \nonumber\\ &&
-\sum_{\rm mesons}^{}
\frac{W_A^+ \left( M,-(2u+\Sigma)/4F \right) -
W_A^+ \left( M,(2s+\Sigma)/4F \right)}{\Sigma + s + u}\ ;
  \\
\label{psi-Aminus-Dt}
\Psi_t (A^{-}) \equiv  &&
\sum_{\rm baryons}^{}
\left[
\frac{Y_A^- (M , -(u+\Sigma))}{s-M^2} +
\frac{Y_A^- (M , -(s+\Sigma))}{u-M^2}
\right]  \nonumber\\ &&
-\sum_{\rm mesons}^{}
\frac{W_A^- \left( M,-(2u+\Sigma)/4F \right) -
W_A^- \left( M,(2s+\Sigma)/4F \right)}{\Sigma + s + u}\ ;
   \\
\label{psi-Bplus-Dt}
\Psi_t (B^{+}) \equiv  &&
\sum_{\rm baryons}^{}
\left[
\frac{Y_B^+ (M , -(u+\Sigma))}{s-M^2}
+ \frac{Y_B^+ (M , -(s+\Sigma))}{u-M^2}
\right]  \nonumber\\ &&
-\sum_{\rm mesons}^{}
\frac{W_B^+ \left( M,-(2u+\Sigma)/4F \right) -
W_B^+ \left( M,(2s+\Sigma)/4F \right)}{\Sigma + s + u}\ ;
  \\
\label{psi-Bminus-Dt}
\Psi_t (B^{-}) \equiv  &&
\sum_{\rm baryons}^{}
\left[
\frac{Y_B^- (M , -(u+\Sigma))}{s-M^2}
-\frac{Y_B^- (M , -(s+\Sigma))}{u-M^2}
\right]  \nonumber\\ &&
-\sum_{\rm mesons}^{}
\frac{W_B^- \left( M , -(2u+\Sigma)/4F \right) -
W_B^- \left( M , (2s+\Sigma)/4F \right)}{\Sigma + s + u}\ .
\end{eqnarray}
\end{widetext}

These functions generate the bootstrap conditions
(\ref{corner-Dt}):
\begin{eqnarray}
&& {\partial}^m_u\, {\partial}^n_s\,
\Psi_t (X_t){\Bigr|}_{u=s=0} = 0,
\nonumber\\&&
(m,n=0,1,\ldots)\, ,\ \ \ \
(X=A^{\pm}, B^{\pm}).
\label{bootstrap-Aplusminus-Bplusminus-Dt}
\end{eqnarray}

It should be stressed once more that the expansion points
$(t,u)=(0,0)$
and
$(s,u)=(0,0)$
in
Eqs.~(\ref{bootstrap-Aplus-Ds}),
(\ref{bootstrap-Aminus-Bplus-Bminus-Ds})
and
(\ref{bootstrap-Aplusminus-Bplusminus-Dt})
are chosen just for convenience. Any other point from the
corresponding domains would be equally acceptable.

As it was already emphasized, the bootstrap constraints restrict the
allowed values of the physical (experimentally observable) parameters.
This is true with respect to the constraints of arbitrary level, and
in particular, with respect to tree level ones. Therefore, the direct
comparison of the constraints
(\ref{bootstrap-Aplus-Ds}),
(\ref{bootstrap-Aminus-Bplus-Bminus-Ds}),
(\ref{bootstrap-Aplusminus-Bplusminus-Dt})
with known data is quite allowable. Unfortunately, the modern data on
the resonance spectrum are far from being complete. Nevertheless, in
two subsequent Sections we will show that it is possible to choose
certain subsystem of constraints under consideration such that the
total contribution from heavy resonances turns out small due to rapid
convergence of the relevant series.

\section{Sum rules for
$\rho$-meson
coupling constants}
\label{sec-rho-couplings}

In this Section we perform the detailed numerical analysis of two
particular bootstrap constraints (sum rules) that connect among
themselves the parameters of baryon and meson spectra. This allows us
to demonstrate an astonishing balance between the numerical values of
two
$\rho NN$
physical coupling constants
$G^T_{NN\rho}$
and
$G^V_{NN\rho}$
and (also physical) parameters of the baryon spectrum.

The quantities
$G^T_{NN\rho}$
and
$G^V_{NN\rho}$
are defined (see
\cite{Nagels})
as coupling constants in the effective Hamiltonian (below
$\sigma_{\mu\nu} \equiv -\frac{i}{4}[\gamma_\mu,\gamma_\nu]_-$)
\begin{eqnarray}
H^{NN\rho}_{\rm eff} =&& -\overline{N}\;
\left[
G^V_{NN\rho}\; \gamma_\mu\; \bo{\rho}^\mu \right.
\nonumber\\ &&
 -\left. G^T_{NN\rho}\frac{\sigma_{\mu\nu}}{4 m}\;
\left(\partial^\mu\bo{\rho}^\nu -\partial^\nu\bo{\rho}^\mu\right)
\right]\;
\frac{1}{2}\; \bo{\sigma}\; N\; . \nonumber\\
\end{eqnarray}
Our constants
$g_{NN\rho}^{(1)}$
and
$g_{NN\rho}^{(2)}$
introduced in
(\ref{H-VNN})
are related to
$G^V_{NN\rho}$
and
$G^T_{NN\rho}$
as follows:
\begin{eqnarray*}
&& g_{NN\rho}^{(1)}  \equiv
\frac{1}{2m}\; G^T_{NN\rho}\; ,
\nonumber \\ &&
g_{NN\rho}^{(2)} \equiv
\frac{G^V_{NN\rho} - G^T_{NN\rho}}{2}\; ,
\end{eqnarray*}
and
$G_{\pi \pi \rho}$
defined in
\cite{Nagels}
differs from our one by the factor of
$2$:
\[
g_{\rho\pi\pi} \equiv = 2 G_{\pi\pi\rho}\; .
\]

The existing experimental data (see
\cite{Nagels})
give:
\begin{eqnarray}
&& \frac{G^T_{NN\rho}}{G^V_{NN\rho}} \approx 6.1 \pm 0.6\, ,\ \ \
\frac{G_{\pi \pi \rho}G^V_{NN\rho}}{4\pi} \approx 2.4 \pm 0.4\, ,\ \ \
\nonumber\\ &&
\ G_{\pi \pi \rho} \approx 6.0\ .
\label{gt/gvexp}
\end{eqnarray}

Let us now take
$\Psi_s (B^-)$
from
(\ref{psi-Bminus-Ds}),
$\Psi_t (A^{-})$
from
(\ref{psi-Aminus-Dt}),
and consider the forms
(\ref{bootstrap-Aminus-Bplus-Bminus-Ds}),
(\ref{bootstrap-Aplusminus-Bplusminus-Dt})
at
$m,n=0$
(i.e. without derivatives). This yields two numerical relations:
\begin{eqnarray}
&& \sum_{\rm baryons}^{}
\left\{ \frac{Y^-_B(M , -\Sigma) - Y^-_B(M, 0)}{\Sigma} -
\frac{Y^-_B(M , 0)}{M^2} \right\} =  \nonumber \\&&
-\, \sum_{\rm mesons \atop {\rm with}\,\, I=1}
\frac{W^-_B(M , \Sigma/4F)}{M^2}\ ;
\label{PsiB-} \\
&& \sum_{\rm baryons}^{} \frac{Y^-_A(M ,-\Sigma)}{M^2} =
\sum_{\rm mesons \atop {\rm with}\,\, I=1}^{}
\frac{W^-_A(M , \Sigma/4F)}{\Sigma}\; ,
\label{PsiA-}
\end{eqnarray}
which can be compared with known data on resonance parameters. The
$\pi N$-resonances
with spin
$J=~l~+~1/2,\ (l=0,1,2,\ldots)$
and isospin
$I=1/2,\ 3/2$,
as well as the isovector
$\pi\pi$-resonances
with spin
$J=1,3,\ldots$
contribute to these equations. It should be probably stressed again
that the summation is performed in order of increasing mass regardless
of the other quantum numbers of contributing resonances. As long as we
can rely on existing experimental values of contributing parameters,
both series above converge very fast. Actually, only four baryons
($N(940),\ N(1440), \, N(1520)$
and
$\Delta (1232)$)
and one meson
($\rho (770)$)
provide significant contributions. This allows one to neglect the
heavier resonances when performing the numerical verification of sum
rules under consideration.

Using the relations of
Sec.~\ref{sec-Cauchy}
and the values
(\ref{gt/gvexp})
of three
$\rho$-meson
coupling constants
$G_{\pi \pi \rho}$, $G^V_{NN\rho}$
and
$G^T_{NN\rho}$,
one can easily estimate the
$\rho$-meson
contributions to the right sides of
(\ref{PsiB-})
and
(\ref{PsiA-}).
The values of baryon resonance parameters given in
Appendix~\ref{app-resonance-list}
allow one to do the same with respect to the left sides. In the case
when we take account of all resonances with
$ M_R \le 1.52$ GeV
in the baryon sector this results in the following numerical relations:
\begin{eqnarray*}
&& {\rm Eq.}~(\ref{PsiB-}):\ \ \ \
324.7 \pm 24 \approx 254 \pm 85;\
\nonumber\\&&
{\rm Eq.}~(\ref{PsiA-}):\ \ \ \
42 \pm 6 \approx 50 \pm 12.5.
\end{eqnarray*}
The uncertainties of right sides should not be taken too seriously ---
these numbers are just indicative
(see
\cite{Nagels}
and references therein). In contrast, the left sides are estimated in
accordance with the numbers given in
Appendix~\ref{app-resonance-list}.
As we just mentioned, the contributions from heavier baryon resonances
turn out to be small, which gives a hope that the above series
converges rapidly enough and eventual (yet unknown) heavy resonances
will not change the sum considerably. This point is graphically
illustrated in the next Section. One may see that both sum rules
(\ref{PsiB-})
and
(\ref{PsiA-})
are quite consistent with known data on the resonance spectrum, as
long as the only resonances taken into account are baryons with masses
$M \le 1.52$ GeV
and the meson
$\rho(770)$.
This coincides well with the so-called
{\em local cancellation hypothesis}
(see the series of papers
\cite{Schechter}).

What happens when the contributions from heavier resonances are
included? In fact, the left side of sum rule
(\ref{PsiA-})
remains almost unchanged until the baryon resonance
$\Delta(1950)$
$(J=7/2;\, \mathcal{N}=-1)$
is taken into account. Its contribution slightly destroys the balance.
As to the sum rule
(\ref{PsiB-}),
the same phenomenon exhibits itself even earlier: already the
contribution from
$N(1680)$
$(J=5/2;\, \mathcal{N}=+1)$
results in small imbalance. In both cases the explanation is quite
obvious: to treat the series correctly (in order of increasing mass)
one needs to take account of the contributions from heavier meson
resonances (in particular, from
$\rho(1450)$)
in the right sides. Unfortunately, the modern experimental data on
the relevant parameters of those resonances are insufficiently
complete to make this possible.

We shall conclude that both bootstrap constraints (sum rules)
(\ref{PsiB-})
and
(\ref{PsiA-})
look quite reasonable from the modern experimental viewpoint. In
particular, one can consider
$G^V_{NN\rho}$
and
$G^T_{NN\rho}$
as unknown parameters and get estimates for them from
Eqs.~(\ref{PsiB-}), (\ref{PsiA-})
(see, e.g.,
\cite{MENU}).
What is interesting to note, is that these constraints possess a
supersymmetric feature --- they connect among themselves the
properties of meson and baryon spectra.


\section{Numerical testing of sum rules for
$\pi N$
spectrum parameters}

In this Section we perform a more detailed numerical testing of the
second kind bootstrap conditions (sum rules)
(\ref{bootstrap-Aplus-Ds}),
(\ref{bootstrap-Aminus-Bplus-Bminus-Ds})
and
(\ref{bootstrap-Aplusminus-Bplusminus-Dt})
for the parameters of pion-nucleon resonance spectrum.

We stress once more that in our effective scattering theory approach
the system of bootstrap conditions (irrelevantly to their level) gives
a set of constraints for the
{\em physical values}
of spectrum parameters. That is why the numerical testing of the tree
level constraints is highly demanding: it allows one to check whether
our scheme is applicable for realistic scattering processes.

The numerical testing of constraints in the toy bootstrap model
(Lovelace string-like amplitude) was successfully carried out in
\cite{POMI}.
In the case of pion-nucleon scattering the situation is a bit more
complicated, since experimental information on resonances is
incomplete --- only the initial part of spectrum is relatively well
established. This may cause certain problems because it is not known
in advance whether a given sum rule converges sufficiently rapidly.
Besides, the physical spectrum, as a rule, contains some poorly
established resonances. The corresponding contributions to sum rules
cannot be estimated with sufficient accuracy.

Nevertheless, as shown below, many of bootstrap constrains for the
parameters of
$\pi N$
spectrum seem to converge sufficiently rapidly. In practice they are
saturated by several lightest well established resonances; the heavier
ones just add small corrections.

\begin{figure}
\includegraphics[height=5.15cm]{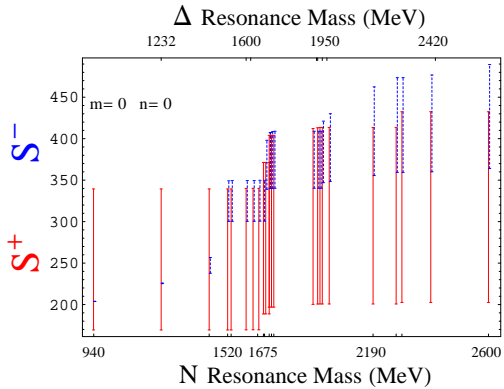}
\caption{\label{Fig_PsiSBminus_m0_n0}
Numerical testing of sum rule following from
the bootstrap condition
(\ref{bootstrap-Aminus-Bplus-Bminus-Ds})
for
$B^-$
in
$D_s$
at
$m=n=0$. }
\end{figure}

\begin{figure*}
\includegraphics[height=5.15cm]{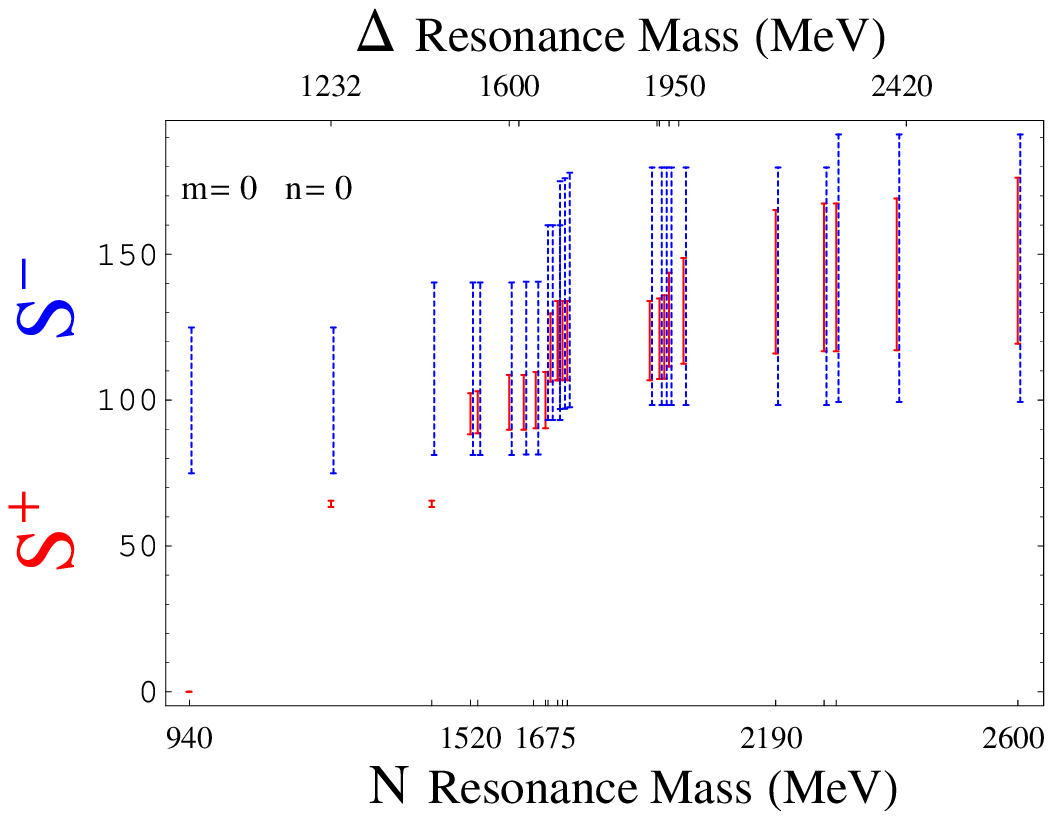}
\includegraphics[height=5.15cm]{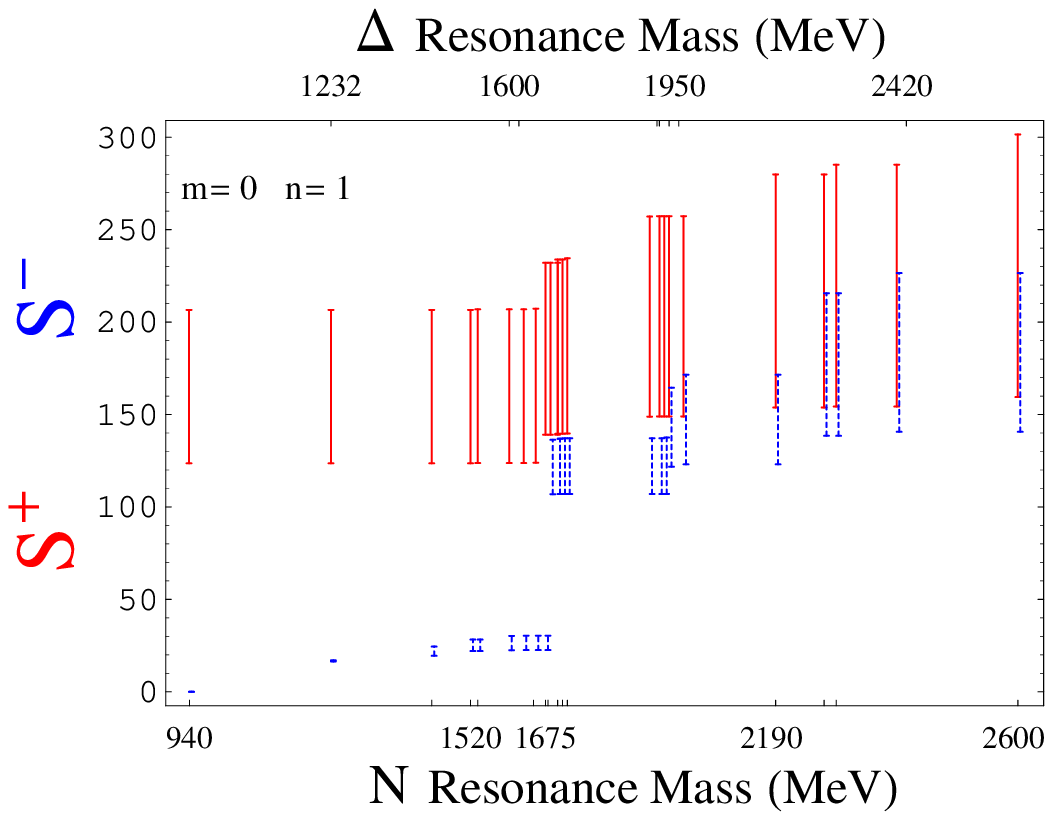}

\includegraphics[height=5.15cm]{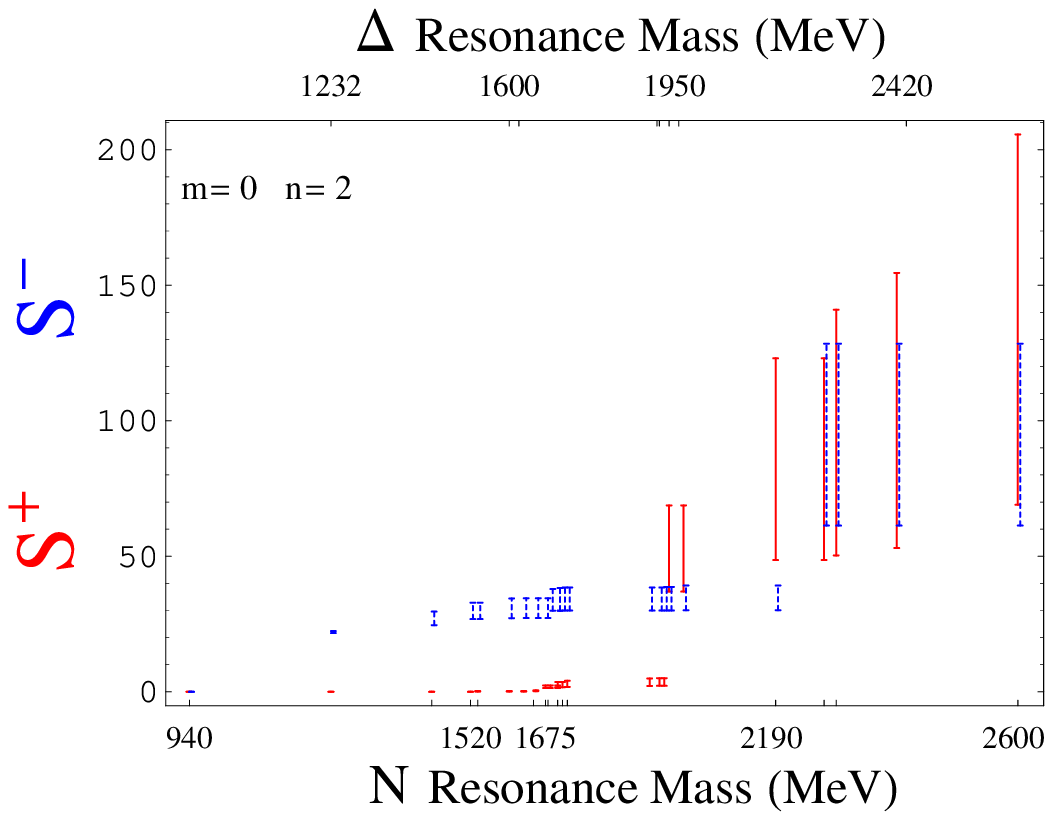}
\includegraphics[height=5.15cm]{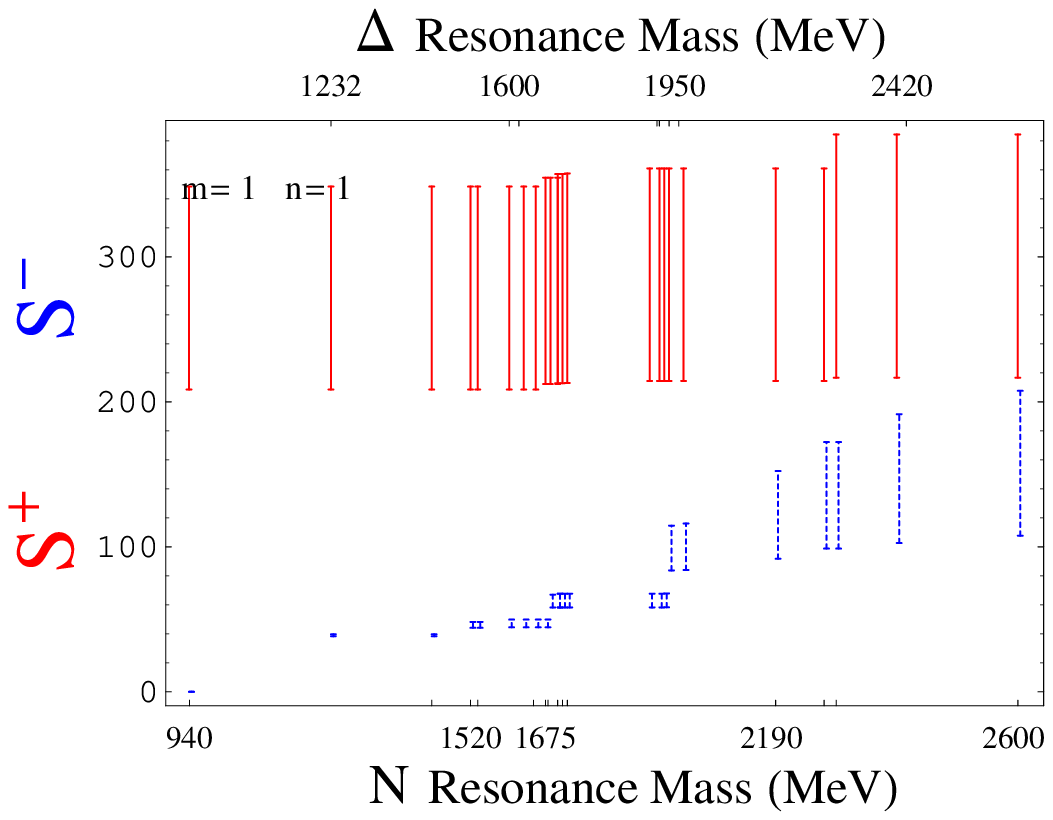}

\caption{\label{Fig_PsiSAminus}Numerical testing of sum rules
following from the bootstrap condition
(\ref{bootstrap-Aminus-Bplus-Bminus-Ds})
for
$A^-$
in
$D_s$
for different values of
$m$
and
$n$.}
\end{figure*}

\begin{figure*}
\includegraphics[height=5.15cm]{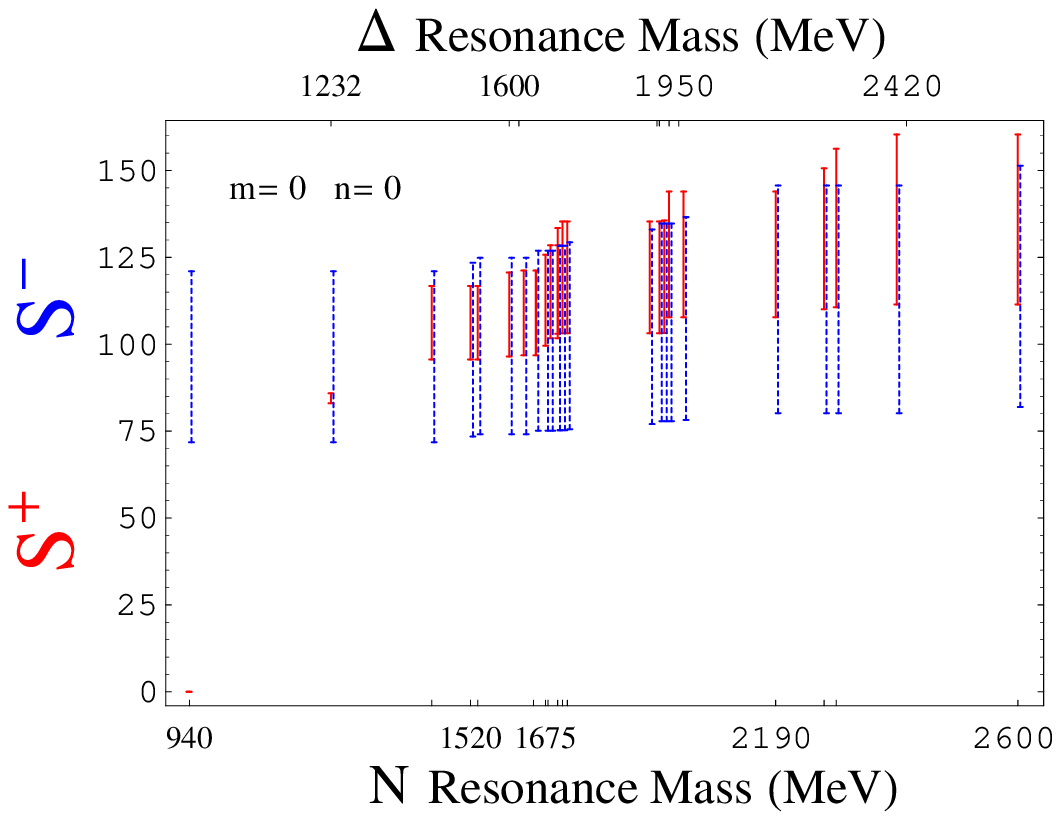}
\includegraphics[height=5.15cm]{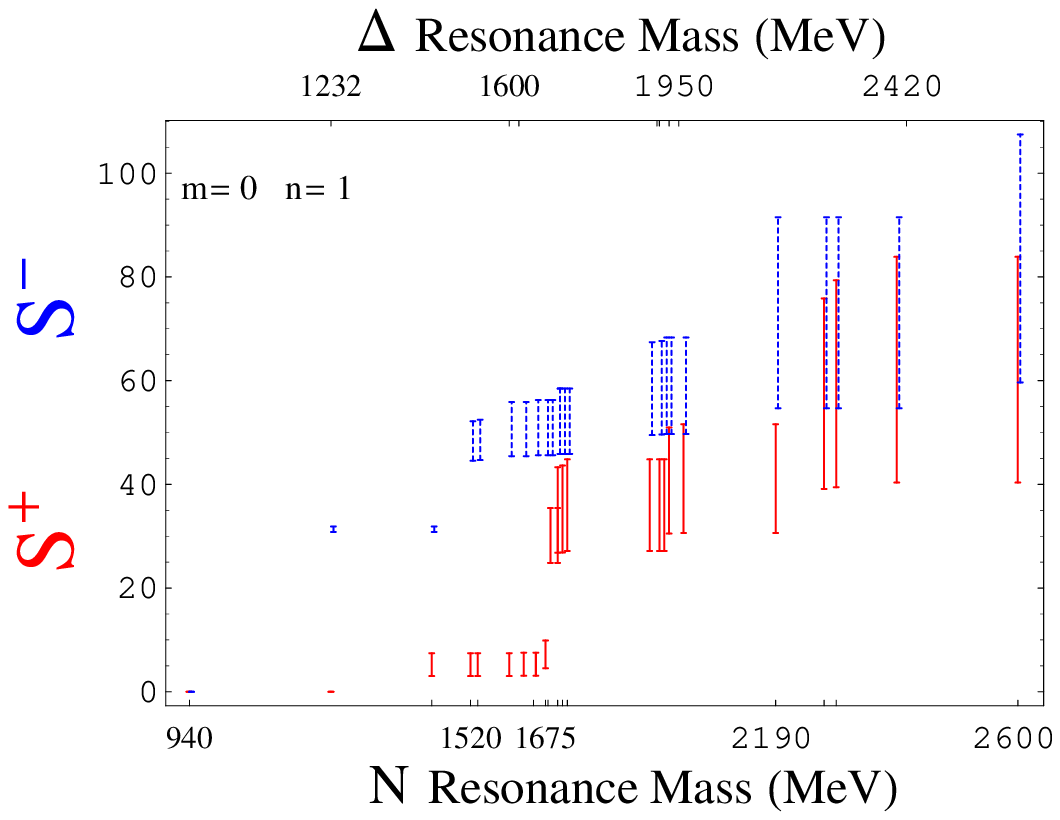}

\includegraphics[height=5.15cm]{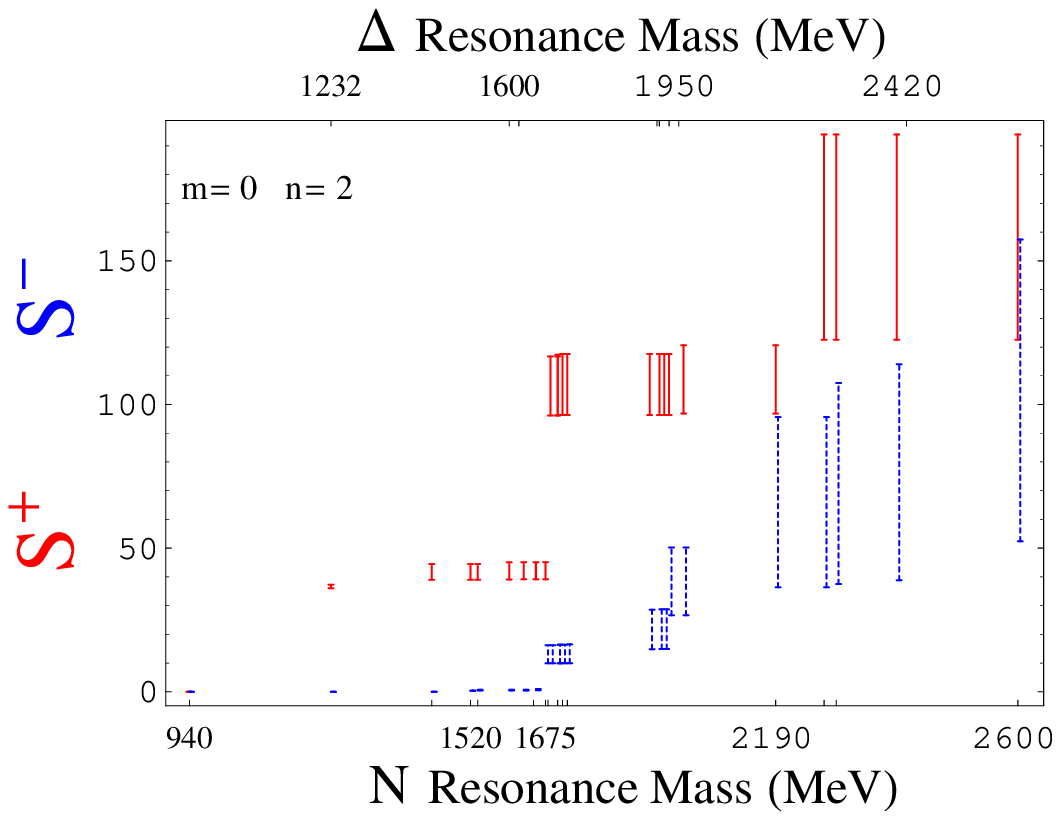}
\includegraphics[height=5.15cm]{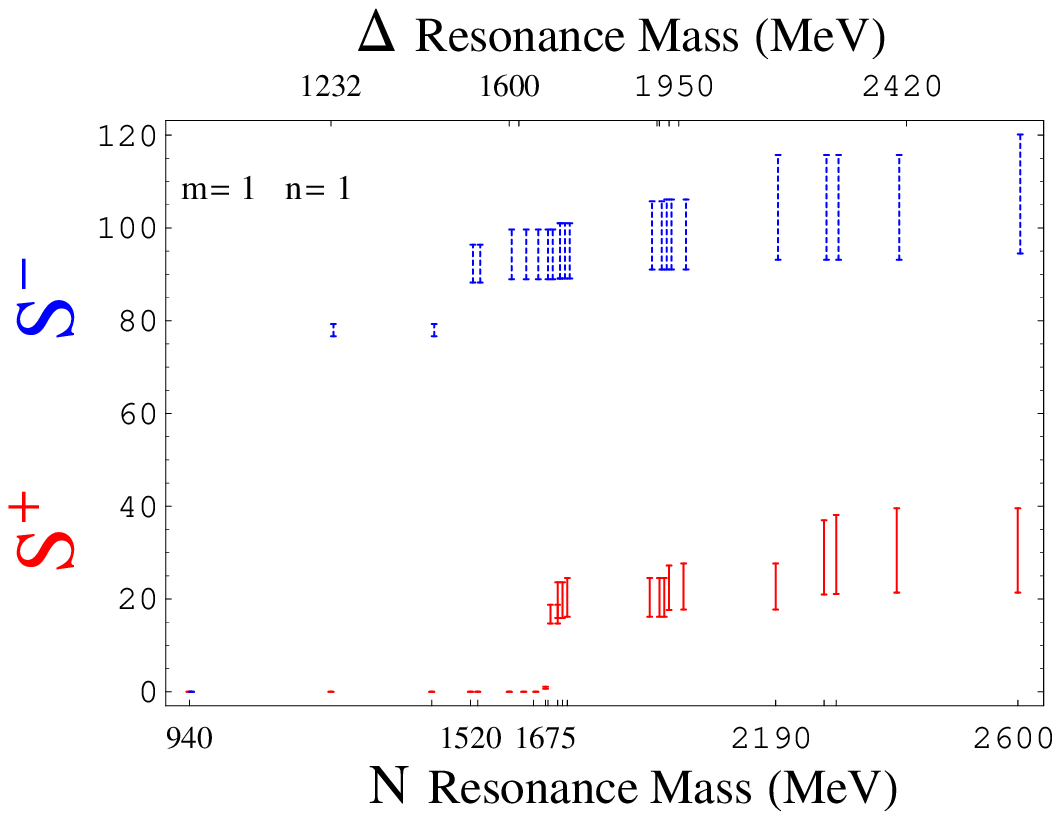}

\caption{\label{Fig_PsiTAminus}
Numerical tests of sum rules following from bootstrap condition
(\ref{bootstrap-Aplusminus-Bplusminus-Dt})
for the amplitude
$A^-$
in
$D_t$
at different values of
$m$
and
$n$.
}
\end{figure*}

To demonstrate the saturation we consider the balance of a given sum
rule as a function of the heaviest resonance mass taken into account.
For this we introduce partial sums of positive and negative
contributions:
$S^+(M_R)$
and
$S^-(M_R)$,
respectively. For example, consider the sum rules which follow from
the constrains
(\ref{bootstrap-Aminus-Bplus-Bminus-Ds})
for the invariant amplitude
$A^-$
in
$D_s$
(the relevant generating function
$\Psi_s(A^-)$
is given in
(\ref{psi-Aminus-Ds})).
For particular
$m$
and
$n$
we define:
\begin{eqnarray*}
S^+(M)= && \sum_{{R_s \, R_t \, R_u,} \atop  M_R \le M}
\left. \frac{\partial^{m+n} \psi_s(A^-)}{\partial t^m \partial u^n}
 \right._{t=0 \atop u=0},
 \nonumber \\&&
{\rm where\; every\; term}\ \ \ \
\left. \frac{\partial^{m+n}
 \psi_s(A^-)}{\partial t^m \partial u^n}
\right._{t=0 \atop u=0} \ge 0;
\end{eqnarray*}
\begin{eqnarray*}
S^-(M )= && \sum_{{R_s \, R_t \, R_u,} \atop  M_R \le M}
\left| \left. \frac{\partial^{m+n}
 \psi_s(A^-)}{\partial t^m \partial u^n}
\right._{t=0 \atop u=0} \right|,
 \nonumber \\&&
{\rm where\; every\; term}\ \ \ \
\left.
 \frac{\partial^{m+n} \psi_s(A^-)}{\partial t^m \partial u^n}
\right._{t=0 \atop u=0}<0.
\end{eqnarray*}
Here
$\psi_s(A^-)$
is an individual resonance contribution to the generating function
$\Psi_s(A^-)$.
These notations allow one to present the sum rule under consideration
as follows:
$$
S^+(M)+ \ldots = S^-(M) + \ldots\, ,
$$
where ellipses stand for the relevant contributions of resonances with
$M_R > M$.
Obviously, when
$S^+ \approx S^-$
the sum rule can be considered as a well saturated one. On Figures
\ref{Fig_PsiSBminus_m0_n0},
\ref{Fig_PsiSAminus},
\ref{Fig_PsiTAminus}
we present several examples of the dependence of
$S^+$
and
$S^-$
on the mass of heaviest baryon resonance taken into account.
The error bars for
$S^+$
and
$S^-$
originate mainly from the uncertainties of decay widths (or, the same,
from those of triple
$\pi NR$
couplings). To make the domains of intersection of error bars better
visible on our Figures
\ref{Fig_PsiSBminus_m0_n0} -- \ref{Fig_PsiTAminus}
the error bars corresponding to
$S^-$
are shifted by 5 MeV to the right from the resonance position.



Some difficulties may arise if a sum rule gets significant
contribution from the meson sector, because the spectrum of heavy
non-strange mesons is known with much less precision than that
of baryon resonances. In this case it makes sense to choose for
numerical testing those sum rules which may only acquire contributions
from meson resonances with
$I=1$.
In many cases the contribution of well established
$\rho(770)$
meson turns out to be the dominant one. Two sum rules of this kind
have been discussed in the previous Section. On the
Figure~\ref{Fig_PsiSBminus_m0_n0}
it is graphically shown the process of saturation of the bootstrap
condition
(\ref{bootstrap-Aminus-Bplus-Bminus-Ds})
at
$m=n=0$.

In this case the positive contribution of
$\rho(770)$
meson is compensated by the contributions from nucleon,
$\Delta(1232)$
and
$N(1440)$.
The contributions due to heavier baryon resonances seem to slightly
disturb the balance. As noted above, this can be explained as a result
of our poor knowledge of the contributions from baryons with
$M > 2$ GeV
and from heavier mesons (say,
$\rho(1450)$).

Now let us consider the sum rules that follow from the bootstrap
constrains
(\ref{bootstrap-Aplus-Ds}),
(\ref{bootstrap-Aminus-Bplus-Bminus-Ds}),
(\ref{bootstrap-Aplusminus-Bplusminus-Dt})
with derivatives (i.e. $m,n \ne 0$).
It is necessary to stress that the saturation of such sum rules
requires attracting the more detailed information on spectrum because
of the following reasons:
\begin{itemize}
\item
The influence of heavy resonances with high spin becomes relatively
more important. This is just because the differentiation kills the
contributions of well established low spin resonances.
\item
The sum rules that arise from bootstrap conditions with derivatives in
some cases converge slowly. This is explained by the fact that the
resonances closest to the domain
$D_x$
under consideration may give significant contribution due to the
presence of small parameter in the denominator. To compensate
gradually their contributions one needs to take account of a large
number of cross channel resonances. Such a situation was encountered
during the numerical testing of sum rules in the toy bootstrap model
for the Lovelace amplitude (see
\cite{POMI}).
\end{itemize}
However, it turns out possible to point out a series of the bootstrap
constrains with derivatives that are reasonably well saturated with
known experimental data. As an example of such sum rules let us
consider several bootstrap conditions
(\ref{bootstrap-Aminus-Bplus-Bminus-Ds})
for the invariant amplitude
$A^-$
at the domain
$D_s$.
The result of saturation of these sum rules for different values of
$m$
and
$n$
is presented on Figure
~\ref{Fig_PsiSAminus}.
Note that these sum rules acquire contributions from
$I=1; \;  J^P=1^-, \,  3^-,...$
meson resonances while we only take into account that of
$\rho(770)$.

As a second example we have chosen a series of purely baryon sum rules
that follow from bootstrap constrains for the same invariant amplitude
$A^-$
in another intersection domain, namely, in
$D_t$.
The results are presented on
Fig.~\ref{Fig_PsiTAminus}.
These sum rules (except that corresponding to
$m=n=1$)
can be considered as reasonably well saturated with known experimental
data. It is interesting to notice that the similar situation was also
encountered in the
``toy bootstrap model''
for Veneziano string amplitude
\cite{POMI}.
In certain sum rules for resonance parameters of the string
amplitude it was sufficient to take into account the contribution
of a relatively small number of first poles to saturate it with high
precision. At the same time, in some other sum rules it was necessary
to take into account the contribution of considerable number of poles
to compensate the `accidentally large' contribution coming from
several first poles. A more detailed information on resonance spectrum
is required to saturate slowly converging sum rules like
(\ref{bootstrap-Aminus-Bplus-Bminus-Ds})
with
$m=n=1$.

Similar series of well saturating sum rules can also be derived from
the bootstrap conditions for other invariant amplitudes
($A^+, \, B^\pm$)
in the domains
$D_s$
and
$D_t$.
This is unlikely to be just an accidental luck. Instead, it gives
serious arguments that the bootstrap constraints for pion-nucleon
spectrum are supported by modern data. Since these constraints appear
as the necessary consistency conditions in the extended perturbation
scheme, this fact can be regarded as a strong evidence in favor of the
latter one.

It is essential that the sum rules of this kind can be used as a
powerful tool in studying the hadron resonance spectrum. This aspect
will be discussed in more detail in the next paper devoted to the
analysis of bootstrap constraints for the elastic kaon-nucleon
scattering amplitude.

\section{Low-energy coefficients}
\label{sec-LEC}

In this Section we present our estimates for the expansion
coefficients of tree level amplitudes around the cross-symmetric point
($t=0,\, \nu_t=0$)
in
$B_t$.
These results present certain interest because those coefficients
undoubtedly do acquire contributions from the loop graphs.
Nevertheless, as shown below, our estimates based on the tree level
approximation of extended perturbation scheme turn out to be in nice
agreement with the known data. This fact demonstrates that the latter
scheme provides quite reasonable numbers already at tree level and,
hence, may be of interest from the computational point of view.

Introducing the new quantity
\[
C^\pm = A^\pm + \frac{m \nu_t}{4m^2 - t} \tilde{B}^\pm\; ,
\]
(here
$\tilde{B}^{\pm}$
is just
$B^{\pm}$
with the nucleon pole subtracted%
\footnote{At this point we follow the definitions accepted in
\cite{Nagels}.
})
we define the low-energy coefficients (LEC's)
$a^{\pm}_{mn}$, $b^{\pm}_{mn}$,
and
$c^{\pm}_{mn}$
as those in double Taylor series expansions around the cross-symmetric
point
$(t=0,\; \nu_t=0)$:
\begin{eqnarray*}
&& \tilde{B}^+ (t,\nu_t) =
\nu_t \sum_{m,n}^{} b^+_{mn} (\nu_t^2)^m t^n\; ;
\nonumber\\&&
\tilde{B}^- (t,\nu_t) =\ \ \
\sum_{m,n}^{} b^-_{mn} (\nu_t^2)^m t^n\; ;
\nonumber\\&&
A^+ (t,\nu_t) =\ \ \
\sum_{m,n}^{} a^+_{mn} (\nu_t^2)^m t^n\; ;
\nonumber\\&&
A^- (t,\nu_t) =
\nu_t\sum_{m,n}^{} a^-_{mn} (\nu_t^2)^m t^n\; ;
\nonumber \\&&
C^+ (t,\nu_t) =\ \ \
\sum_{m,n}^{} c^+_{mn} (\nu_t^2)^m t^n\; ;
\nonumber\\ &&
C^- (t,\nu_t) =
\nu_t\sum_{m,n}^{} c^-_{mn} (\nu_t^2)^m t^n\; .
\end{eqnarray*}

To get numerical values for these coefficients, we need to re-expand
the Cauchy forms
(\ref{ACauchyBt}), (\ref{CauchyBt})
and
(\ref{alpha-t})
in double power series in
$(t, \nu_t)$.
This is quite admissible because these forms converge uniformly in
whole
$B_t$
and, therefore, near the cross-symmetric point.

Now, using the data
\cite{Nagels}, \cite{PDG}
(see also the Table in
Appendix~\ref{app-resonance-list})
on coupling constants and masses and neglecting the contributions of
resonances with
$M\geq 1.95$ GeV,
one can get the theoretical estimates for these coefficients and
compare them with known numbers
\cite{Nagels},
which follow from independent theoretical processing of experimental
data. The results are collected in six Tables below.
Note that in
\cite{Nagels}
somewhat different definitions of low-energy coefficients are used, so
one needs to perform certain rescaling to compare the results. This is
already done in the Tables
\ref{table-Aplus} -- \ref{table-Cminus}.

When computing the LEC's we have used the data
\cite{Nagels}
and
\cite{PDG}
for the resonance parameters (listed in
Appendix~\ref{app-resonance-list});
the estimated errors correspond to maximal and minimal values of the
quantity under consideration. In order to save space we use the
following shortened form of number recording:
$X^n \equiv X \times 10^n$.

In the first two lines of Tables
\ref{table-Aminus}
$\div$
\ref{table-Bminus}
and
\ref{table-Cminus}
(three lines in Tables
\ref{table-Aplus}
and
\ref{table-Cplus})
we also show the most significant individual contributions --- those
coming from
$\Delta(1232)$
and
$N(1440)$
(and from the scalar
$\sigma$
meson in the Tables
\ref{table-Aplus}
and
\ref{table-Cplus}).
The line
{\em Full set}
shows the results of summing over contributions from all the
resonances listed in
Appendix~\ref{app-resonance-list}.
The results of independent theoretical analysis of
experimental data (the lines
{\em Data}
in Tables
\ref{table-Bplus} -- \ref{table-Cminus})
are taken from
\cite{Nagels}.
The lines
{\em Data}
are absent in Tables
\ref{table-Aplus}, \ref{table-Aminus}
because the corresponding numbers are not available in
\cite{Nagels}.
It should be kept in mind that the errors shown in the lines
{\em Data}
are just indicative. The reason is that the corresponding numbers
strongly depend upon various theoretical suggestions (like, say, the
value of
$S$-wave
pion-pion scattering length with isospin
$I=0$;
see
\cite{Nagels})
used as the theoretical input in the process of data analysis.
Clearly, it would make no sense to show the error bars in the lines
which correspond to
$\sigma$-meson
contributions (Tables
\ref{table-Aplus}
and
\ref{table-Cplus}).

\begin{table*}
\caption{\label{table-Aplus} Tree level low energy coefficients
$a^+_{mn}$
$(m,n=0,1,2)$.}
\begin{ruledtabular}
\begin{tabular}{|l|l|l|l|l|l|l|l|l|l|}
\hline
{\tiny Resonance}
&$  a^+_{00}   $&$  a^+_{01}   $&$  a^+_{02}   $
&$  a^+_{10}   $&$  a^+_{11}   $&$  a^+_{12}   $
&$  a^+_{20}   $&$  a^+_{21}   $&$  a^+_{22}   $
\\ \hline
${\scriptstyle \sigma (650)}$
&$ +19         $&$ +0.94       $&$ +5.20^{-2} $
&$             $&$             $&$            $
&$             $&$             $&$            $
\\
&$             $&$             $&$            $
&$             $&$             $&$            $
&$             $&$             $&$            $
\\ \hline
${\scriptstyle \Delta (1232)}$
&$ +2.74       $&$ +7.19^{-1}  $&$  -1.66^{-2} $
&$ +6.36       $&$ -1.27^{-2}  $&$  -4.22^{-3} $
&$ +1.16       $&$ -3.91^{-2}  $&$  +1.83^{-4} $
\\
&$\pm 4.6^{-2} $&$\pm 1.2^{-2} $&$\pm 2.8^{-4} $
&$\pm 1.1^{-1} $&$\pm 2.1^{-4} $&$\pm 7.2^{-5} $
&$\pm 2.0^{-2} $&$\pm 6.6^{-4} $&$\pm 3.1^{-6} $
\\ \hline
${\scriptstyle N(1440)} $
&$ -3.86       $&$ +4.50^{-2}  $&$  -3.76^{-4} $
&$ -2.71^{-1}  $&$ +6.79^{-3}  $&$  -1.13^{-4} $
&$ -1.36^{-2}  $&$ +5.69^{-4}  $&$  -1.43^{-5} $
\\
&$\pm 1.6      $&$\pm 1.9^{-2} $&$\pm 1.6^{-4} $
&$\pm 1.1^{-1} $&$\pm 2.9^{-3} $&$\pm 4.8^{-6} $
&$\pm 5.7^{-3} $&$\pm 2.4^{-4} $&$\pm 6.0^{-6} $
\\ \hline
{\tiny Full set}
&$ +23.1       $&$ +1.63       $&$ +3.50^{-2}  $
&$ +6.03       $&$ -1.02^{-2}  $&$ -4.23^{-3}  $
&$ +1.14       $&$ -3.86^{-2}  $&$ +1.73^{-4}  $
\\
&$\pm 6.6      $&$\pm 1.2^{-1} $&$\pm 1.7^{-3} $
&$\pm 4.4^{-1} $&$\pm 8.9^{-3} $&$\pm 1.9^{-4} $
&$\pm 2.9^{-2} $&$\pm 9.7^{-4} $&$\pm 1.0^{-5} $
\\ \hline
\end{tabular}
\end{ruledtabular}
\end{table*}

\begin{table*}
\caption{\label{table-Aminus} Tree level low energy coefficients
$a^-_{mn}$
$(m,n=0,1,2)$.}
\begin{ruledtabular}
\begin{tabular}{|l|l|l|l|l|l|l|l|l|l|}
\hline
{\tiny Resonance}
&$  a^-_{00}   $&$  a^-_{01}   $&$  a^-_{02}   $
&$  a^-_{10}   $&$  a^-_{11}   $&$  a^-_{12}   $
&$  a^-_{20}   $&$  a^-_{21}   $&$  a^-_{22}   $
\\ \hline
${\scriptstyle \Delta (1232)}$
&$ -7.46       $&$ -1.04^{-1}  $&$  +5.18^{-3} $
&$ -1.36       $&$ +2.43^{-2}  $&$  +5.14^{-4} $
&$ -2.47^{-1}  $&$ +1.22^{-2}  $&$  -2.33^{-4} $
\\
&$\pm 1.3^{-1} $&$\pm 1.6^{-3} $&$\pm 8.8^{-5} $
&$\pm 2.3^{-2} $&$\pm 4.1^{-4} $&$\pm 8.7^{-6} $
&$\pm 4.2^{-3} $&$\pm 2.1^{-4} $&$\pm 4.0^{-6} $
\\ \hline
${\scriptstyle N(1440)} $
&$ -1.21       $&$ +2.02^{-2}  $&$  -2.53^{-4} $
&$ -6.01^{-2}  $&$ +2.03^{-3}  $&$  -4.24^{-5} $
&$ -3.05^{-3}  $&$ +1.53^{-4}  $&$  -4.48^{-6} $
\\
&$\pm 5.1^{-1} $&$\pm 8.5^{-3} $&$\pm 1.1^{-4} $
&$\pm 2.6^{-2} $&$\pm 8.5^{-4} $&$\pm 1.8^{-5} $
&$\pm 1.3^{-3} $&$\pm 6.4^{-5} $&$\pm 1.9^{-6} $
\\ \hline
{\tiny Full set}
&$ -10.5       $&$ -1.80^{-1}  $&$ +4.24^{-3}  $
&$ -1.45       $&$ +2.53^{-2}  $&$ +4.86^{-4}  $
&$ -2.51^{-1}  $&$ +1.24^{-2}  $&$ -2.38^{-4}  $
\\
&$\pm 2.0      $&$\pm 6.4^{-2} $&$\pm 8.6^{-4} $
&$\pm 6.8^{-2} $&$\pm 1.7^{-3} $&$\pm 3.3^{-5} $
&$\pm 5.9^{-3} $&$\pm 2.8^{-4} $&$\pm 6.1^{-6} $
\\ \hline
\end{tabular}
\end{ruledtabular}
\end{table*}

\begin{table*}
\caption{\label{table-Bplus} Tree level low energy coefficients
$b^+_{mn}$
$(m,n=0,1,2)$.}
\begin{ruledtabular}
\begin{tabular}{|l|l|l|l|l|l|l|l|l|l|}
\hline
{\tiny Resonance}
&$  b^+_{00}   $&$  b^+_{01}   $&$  b^+_{02}   $
&$  b^+_{10}   $&$  b^+_{11}   $&$  b^+_{12}   $
&$  b^+_{20}   $&$  b^+_{21}   $&$  b^+_{22}   $
\\ \hline
${\scriptstyle \Delta (1232)}$
&$ -5.20       $&$ +2.09^{-1}  $&$  -5.34^{-3} $
&$ -9.45^{-1}  $&$ +6.81^{-2}  $&$  -2.90^{-3} $
&$ -1.72^{-1}  $&$ +1.79^{-2}  $&$  -1.05^{-3} $
\\
&$\pm 8.8^{-2} $&$\pm 3.5^{-3} $&$\pm 9.0^{-5} $
&$\pm 1.6^{-2} $&$\pm 1.2^{-3} $&$\pm 4.9^{-5} $
&$\pm 2.9^{-3} $&$\pm 3.0^{-4} $&$\pm 1.8^{-5} $
\\ \hline
${\scriptstyle N(1440)} $
&$ +3.37^{-1}  $&$ -5.64^{-3}  $&$  +7.07^{-5} $
&$ +1.70^{-2}  $&$ -5.67^{-4}  $&$  +1.19^{-5} $
&$ +8.53^{-4}  $&$ -4.28^{-5}  $&$  +1.25^{-6} $
\\
&$\pm 1.4^{-1} $&$\pm 2.4^{-3} $&$\pm 3.0^{-5} $
&$\pm 7.1^{-3} $&$\pm 2.4^{-4} $&$\pm 5.0^{-6} $
&$\pm 3.6^{-4} $&$\pm 1.8^{-5} $&$\pm 5.3^{-7} $
\\ \hline
{\tiny Full set}
&$ -4.64       $&$ +2.19^{-1}  $&$ -5.25^{-3}  $
&$ -9.22^{-1}  $&$ +6.78^{-2}  $&$ -2.89^{-3}  $
&$ -1.71^{-1}  $&$ +1.78^{-2}  $&$ -1.05^{-3}  $
\\
&$\pm 4.3^{-1} $&$\pm 1.2^{-2} $&$\pm 1.9^{-4} $
&$\pm 2.7^{-2} $&$\pm 1.5^{-3} $&$\pm 5.6^{-5} $
&$\pm 3.3^{-3} $&$\pm 3.2^{-4} $&$\pm 1.8^{-5} $
\\ \hline\hline
{\tiny Data}
&$ -3.50       $&$ +2.50^{-1}  $&$ -1.00^{-2}  $
&$ +9.6^{-2}   $&$ +4.80^{-2}  $&$ -1.00^{-2}  $
&$ -3.10^{-1}  $&$ +4.80^{-2}  $&$ -9.00^{-3}  $
\\
&$\pm 1.1^{-1} $&$\pm 1.1^{-1} $&$\pm 5.0^{-3} $
&$\pm 2.0^{-2} $&$\pm 4.7^{-2} $&$\pm 2.0^{-3} $
&$\pm 5.0^{-2} $&$\pm 4.7^{-2} $&$\pm 3.0^{-3} $
\\ \hline
\end{tabular}
\end{ruledtabular}
\end{table*}

\begin{table*}
\caption{\label{table-Bminus} Tree level low energy coefficients
$b^-_{mn}$
$(m,n=0,1,2)$.}
\begin{ruledtabular}
\begin{tabular}{|l|l|l|l|l|l|l|l|l|l|}
\hline
{\tiny Resonance}
&$  b^-_{00}   $&$  b^-_{01}   $&$  b^-_{02}   $
&$  b^-_{10}   $&$  b^-_{11}   $&$  b^-_{12}   $
&$  b^-_{20}   $&$  b^-_{21}   $&$  b^-_{22}   $
\\ \hline
${\scriptstyle \Delta (1232)}$
&$ +6.09       $&$ -1.48^{-1}  $&$  +2.36^{-3} $
&$ +1.11       $&$ -6.22^{-2}  $&$  +2.13^{-3} $
&$ +2.02^{-1}  $&$ -1.78^{-2}  $&$  +9.00^{-4} $
\\
&$\pm 1.0^{-1} $&$\pm 2.5^{-3} $&$\pm 4.0^{-5} $
&$\pm 1.9^{-2} $&$\pm 1.1^{-3} $&$\pm 3.6^{-5} $
&$\pm 3.4^{-3} $&$\pm 3.0^{-4} $&$\pm 1.5^{-5} $
\\ \hline
${\scriptstyle N(1440)} $
&$ +1.50       $&$ -1.26^{-2}  $&$  +1.05^{-4} $
&$ +7.56^{-2}  $&$ -1.90^{-3}  $&$  +3.17^{-5} $
&$ +3.80^{-3}  $&$ -1.59^{-4}  $&$  +3.99^{-6} $
\\
&$\pm 6.3^{-1} $&$\pm 5.3^{-3} $&$\pm 4.4^{-5} $
&$\pm 3.2^{-2} $&$\pm 8.0^{-4} $&$\pm 1.3^{-5} $
&$\pm 1.6^{-3} $&$\pm 6.7^{-5} $&$\pm 1.7^{-6} $
\\ \hline
{\tiny Full set}
&$ +9.55       $&$ -4.47^{-2}  $&$ +3.60^{-3}  $
&$ +1.22       $&$ -6.25^{-2}  $&$ +2.15^{-3}  $
&$ +2.07^{-1}  $&$ -1.79^{-2}  $&$ +9.03^{-4}  $
\\
&$\pm 2.0      $&$\pm 6.0^{-2} $&$\pm 6.7^{-4} $
&$\pm 6.9^{-2} $&$\pm 2.4^{-3} $&$\pm 5.7^{-5} $
&$\pm 5.4^{-3} $&$\pm 3.8^{-4} $&$\pm 1.7^{-5} $
\\ \hline\hline
{\tiny Data}
&$ +8.43       $&$ +2.00^{-1}  $&$ +2.00^{-2}  $
&$ +1.08       $&$ -6.30^{-2}  $&$ +4.00^{-3}  $
&$ +3.10^{-1}  $&$ -3.60^{-2}  $&$ +3.00^{-3}  $
\\
&$\pm 1.2^{-1} $&$\pm 1.2^{-1} $&$\pm 8.0^{-3} $
&$\pm 4.0^{-2} $&$\pm 1.2^{-2} $&$\pm 1.9^{-3} $
&$\pm 4.0^{-2} $&$\pm 2.8^{-2} $&$\pm 1.0^{-3} $
\\ \hline
\end{tabular}
\end{ruledtabular}
\end{table*}


\begin{table*}
\caption{\label{table-Cplus} Tree level low energy coefficients
$c^+_{mn}$
$(m,n=0,1,2)$.}
\begin{ruledtabular}
\begin{tabular}{|l|l|l|l|l|l|l|l|l|l|}
\hline
{\tiny Resonance}
&$  c^+_{00}   $&$  c^+_{01}   $&$  c^+_{02}   $
&$  c^+_{10}   $&$  c^+_{11}   $&$  c^+_{12}   $
&$  c^+_{20}   $&$  c^+_{21}   $&$  c^+_{22}   $
\\ \hline
${\scriptstyle \sigma (650)}$
&$ +19         $&$ +0.94       $&$  +5.20^{-2} $
&$             $&$             $&$             $
&$             $&$             $&$             $
\\
&$             $&$             $&$             $
&$             $&$             $&$             $
&$             $&$             $&$             $
\\ \hline
${\scriptstyle \Delta (1232)}$
&$ +2.74       $&$ +7.18^{-1}  $&$  -1.66^{-2} $
&$ +1.17       $&$ +1.68^{-1}  $&$  -8.56^{-3} $
&$ +2.12^{-1}  $&$ +2.38^{-2}  $&$  -2.37^{-3} $
\\
&$\pm 4.6^{-1} $&$\pm 1.2^{-2} $&$\pm 2.8^{-4} $
&$\pm 9.8^{-2} $&$\pm 1.4^{-3} $&$\pm 4.6^{-5} $
&$\pm 1.8^{-2} $&$\pm 6.4^{-4} $&$\pm 1.5^{-5} $
\\ \hline
${\scriptstyle N(1440)} $
&$ -3.86       $&$ +4.50^{-2}  $&$  -3.76^{-4} $
&$ +6.65^{-2}  $&$ +3.02^{-3}  $&$  -6.37^{-5} $
&$ +3.35^{-3}  $&$ +9.59^{-5}  $&$  -5.04^{-6} $
\\
&$\pm 1.6      $&$\pm 1.9^{-2} $&$\pm 1.6^{-4} $
&$\pm 1.3^{-1} $&$\pm 2.0^{-3} $&$\pm 2.4^{-5} $
&$\pm 6.4^{-3} $&$\pm 1.7^{-4} $&$\pm 3.1^{-6} $
\\ \hline
{\tiny Full set}
&$ +23.1       $&$ +1.63       $&$ +3.50^{-2}  $
&$ +1.39       $&$ +1.83^{-1}  $&$ -8.41^{-3}  $
&$ +2.19^{-1}  $&$ +2.40^{-2}  $&$ -2.37^{-3}  $
\\
&$\pm 6.6      $&$\pm 1.2^{-1} $&$\pm 1.7^{-3} $
&$\pm 4.3^{-1} $&$\pm 7.9^{-3} $&$\pm 1.2^{-4} $
&$\pm 2.8^{-2} $&$\pm 8.6^{-4} $&$\pm 1.9^{-5} $
\\ \hline\hline
{\tiny Data}
&$ +25.6       $&$ +1.18       $&$ +3.55^{-2}  $
&$ +1.18       $&$ +1.53^{-1}  $&$ -1.50^{-2}  $
&$ +2.00^{-1}  $&$ +3.40^{-2}  $&$ -8.00^{-3}  $
\\
&$\pm 5.0^{-1} $&$\pm 5.0^{-2} $&$\pm 7.0^{-3} $
&$\pm 5.0^{-2} $&$\pm 1.7^{-2} $&$\pm 3.0^{-3} $
&$\pm 1.0^{-2} $&$\pm 1.0^{-3} $&$\pm 1.0^{-3} $
\\ \hline
\end{tabular}
\end{ruledtabular}
\end{table*}

\begin{table*}
\caption{\label{table-Cminus} Tree level low energy coefficients
$c^-_{mn}$
$(m,n=0,1,2)$.}
\begin{ruledtabular}
\begin{tabular}{|l|l|l|l|l|l|l|l|l|l|}
\hline
{\tiny Resonance}
&$  c^-_{00}   $&$  c^-_{01}   $&$  c^-_{02}   $
&$  c^-_{10}   $&$  c^-_{11}   $&$  c^-_{12}   $
&$  c^-_{20}   $&$  c^-_{21}   $&$  c^-_{22}   $
\\ \hline
${\scriptstyle \Delta (1232)}$
&$ -1.37       $&$ -2.18^{-1}  $&$  +6.91^{-3} $
&$ -2.49^{-1}  $&$ -3.18^{-2}  $&$  +2.33^{-3} $
&$ -4.52^{-2}  $&$ -4.35^{-3}  $&$  +5.74^{-4} $
\\
&$\pm 1.1^{-1} $&$\pm 1.6^{-3} $&$\pm 3.6^{-5} $
&$\pm 2.1^{-2} $&$\pm 5.2^{-4} $&$\pm 1.3^{-5} $
&$\pm 3.8^{-3} $&$\pm 1.8^{-4} $&$\pm 5.2^{-6} $
\\ \hline
${\scriptstyle N(1440)} $
&$ +2.97^{-1}  $&$ +1.59^{-2}  $&$  -1.71^{-4} $
&$ +1.49^{-2}  $&$ +5.52^{-4}  $&$  -1.89^{-5} $
&$ +7.51^{-4}  $&$ +1.52^{-5}  $&$  -1.26^{-6} $
\\
&$\pm 5.7^{-1} $&$\pm 5.8^{-3} $&$\pm 5.0^{-5} $
&$\pm 2.9^{-2} $&$\pm 6.1^{-4} $&$\pm 9.2^{-6} $
&$\pm 1.4^{-3} $&$\pm 4.7^{-5} $&$\pm 1.0^{-7} $
\\ \hline
{\tiny Full set}
&$ -1.00       $&$ -1.77^{-1}  $&$ +7.80^{-3}  $
&$ -2.27^{-1}  $&$ -3.05^{-2}  $&$ +2.33^{-3}  $
&$ -4.43^{-2}  $&$ -4.31^{-3}  $&$ +5.72^{-4}  $
\\
&$\pm 2.0      $&$\pm 4.4^{-2} $&$\pm 4.7^{-4} $
&$\pm 6.8^{-2} $&$\pm 1.5^{-3} $&$\pm 2.6^{-5} $
&$\pm 5.6^{-3} $&$\pm 2.3^{-4} $&$\pm 6.4^{-6} $
\\ \hline\hline
{\tiny Data}
&$ -5.05^{-1}  $&$ -9.70^{-2}  $&$ +9.00^{-3}  $
&$ -1.63^{-1}  $&$ -3.90^{-2}  $&$ -5.00^{-3}  $
&$ -3.80^{-2}  $&$ -1.30^{-2}  $&$ +3.00^{-3}  $
\\
&$\pm 4.5^{-2} $&$\pm 1.2^{-2} $&$\pm 7.0^{-3} $
&$\pm 7.0^{-3} $&$\pm 5.0^{-3} $&$\pm 2.0^{-3} $
&$\pm 4.0^{-3} $&$\pm 4.0^{-3} $&$\pm 1.0^{-3} $
\\ \hline
\end{tabular}
\end{ruledtabular}
\end{table*}

As clearly seen from these Tables, only two lightest baryon resonances
--- $\Delta(1232)$ and $N(1440)$ ---
provide significant contributions to all the coefficients except
$a^+_{0j}$
and
$c^+_{0j}$.
From
(\ref{ACauchyBt}),
(\ref{CauchyBt})
and
(\ref{alpha-t})
it follows that the meson resonances only contribute to
$\alpha (t)$,
the Table
\ref{table-Aplus}
(as well as
\ref{table-Cplus})
shows that the values of
$a^{+}_{00} \div a^{+}_{02}$
($c^{+}_{00} \div c^{+}_{02}$)
cannot be explained if we neglect the contribution due to famous
light scalar
$\sigma$-meson%
\footnote{This statement remains true with respect to
$a^{+}_{03}$
($c^{+}_{03}$).
}
with the mass parameter
$M_{\sigma} \sim 550 \div 700$ MeV
and
``effective coupling''
(see
\cite{Nagels})
$$
G^0_1 \equiv g_{\sss S\pi \pi} g^{\sss (1)}_{\sss NNS} \sim
50 \div 100\; .
$$

Altogether, these results show that the extended perturbation scheme
provides reasonable values for the low energy coefficients already at
tree level. We emphasize that this is closely connected with the
postulated Regge asymptotic conditions in the hyperlayer
$B_t$.
One can check that, once these conditions are violated, results start
to differ drastically (by several orders!) from those shown in Tables
\ref{table-Aplus} -- \ref{table-Cminus}.
Besides, it turns out that the presence of the light scalar meson is
also essential. Although the scalar mesons do not contribute to the
second kind bootstrap conditions, the necessity of introducing the
corresponding auxiliary fields follows from the data on
$c^+_{0j}$ ($a^+_{0j}$).
The simplest way to explain the values of those coefficients is to
suggest the existence of at least one light scalar meson with
above-specified parameters. It is interesting to note that the similar
situation has revealed itself in the case of pion-kaon elastic
scattering (see
\cite{AVVV1}).


\section{Conclusions}
\label{sec-conclusions}

The numerical analysis of bootstrap constraints for the tree level
amplitude of elastic pion-nucleon scattering shows that both physical
(Regge-like asymptotic behavior) and mathematical (uniformity and
summability principles) postulates, used as the basis for extended
perturbation scheme suggested in the series of papers
\cite{AVVV2} -- \cite{AVVV1},
look quite reasonable. In those cases when experimental data on the
resonance spectrum allow to check the consistency of corresponding sum
rules, the results are satisfactory. It is interesting to note that,
in general, these sum rules possess certain features of supersymmetry
since they connect among themselves the parameters of meson and baryon
spectra. Besides, numerical tests show that our sum rules confirm the
so-called local cancellation hypotheses suggested in the series of
papers
\cite{Schechter}.

Moreover, as follows from the results of
Sec.~\ref{sec-LEC},
already the first term (trees) of the extended Dyson series provides
reasonable numerical values for the low energy coefficients which
certainly acquire contributions from the higher order terms. This
gives us a hope that the latter terms will result just in small
corrections. If so, this would mean that the general philosophy of
quasiparticle method (see
\cite{Quasi})
can be successfully applied to the case of effective scattering theory
of strong interaction.

In subsequent publication we will show that these conclusions hold
also for elastic kaon-nucleon scattering.

\section*{Acknowledgements}
\label{sec-thanks}

We are grateful to V.~A.~Franke, H.~Nielsen, P.~Osland, S.~Paston,
J.~Schechter, A.~Tochin, A.~Vasiliev and M.~Vyazovski for stimulating
discussions. The work was supported in part by INTAS
(project 587, 2000) and by the Russian National Programme (grant RNP
2.1.1.1112). The work by A.~Vereshagin was supported by
L.~Meltzers H\o yskolefond (Studentprosjektstipend, 2004).

\appendix
\section{Contracted projecting operators}
\label{sec-contproj}

In practical calculations in the framework of the effective scattering
theory approach one never needs to use the explicit form of spin sums.
It turns out sufficient to exploit the formalism of so-called
{\em contracted projecting operators}
based on the properties of Rarita-Schwinger wave functions. Although
this formalism has been developed nearly forty years ago in the series
of papers
\cite{WeinQuant}
(see also
\cite{WeinMONO}, \cite{Alfaro}, \cite{Scadron}
and references therein), it is still not widely known. That is why in
this Appendix we present a short summary of corresponding formulae.

Let us first consider the case of a free boson field with the mass
parameter
$M$,
momentum
$p_{\mu}$ ($p^2 = M^2$),
spin
$J=l=1,2,...$
and polarization
$i = -l, \ldots, l$.
The corresponding wave function is described by a symmetric traceless
tensor
$
{\cal E}_{\mu_1 \ldots\mu_l} (i,p)
$
which satisfies the following conditions:
\begin{eqnarray*}
&& {\cal E}_{\ldots\mu_m\ldots\mu_n\ldots}(i,p) =
{\cal E}_{\ldots\mu_n\ldots\mu_m\ldots}(i,p)
\ \ \ \ \text{(symmetry)},
\nonumber\\
\nonumber\\&&
g^{\mu_m\mu_n}
{\cal E}_{\ldots\mu_m\ldots\mu_n \ldots}(i,p)=0
\ \ \ \ \text{(tracelessness)},
\nonumber\\
\nonumber\\&&
p^{\mu_m}
{\cal E}_{\ldots\mu_m\ldots}(i,p)=0
\ \ \ \ \text{(transversality)},
\nonumber\\
\nonumber\\&&
\left[{\cal E}_{\mu_1\ldots\mu_l} (i,p)\right]^\ast
{\cal E}^{\mu_1\ldots\mu_l}(j,p)
= (-1)^l\; \delta_{ij}
\nonumber\\&&
\ \ \ \ \ \ \ \ \ \ \ \ \ \ \ \ \ \
\ \ \ \ \ \ \ \ \ \ \ \ \ \ \ \ \ \
\ \ \ \ \ \ \ \ \ \ \
\text{(normalization)}.
\end{eqnarray*}
The
{\em spin sum} is defined as follows (for
$p^2=M^2$)
\begin{equation}
\Pi^{\mu_1 \ldots \mu_l}_{\nu_1 \ldots \nu_l}(p\, ;l)
\equiv
\sum^{l}_{i=-l} {\cal E}^{\mu_1 \ldots \mu_l}(i,p)
 \left[ {\cal E}_{\nu_1 \ldots \nu_l}(i,p) \right]^\ast \; .
\label{spin-sum-boson}
\end{equation}
Typically, when computing a given
$S$-matrix
element one only needs to know the so-called
{\em contracted projector}
\begin{equation}
{\cal P}^{(l)} (k,k',p) \equiv
k^{\nu_1} \ldots k^{\nu_l}
\Pi_{\nu_1 \ldots \nu_l}^{\mu_1 \ldots \mu_l}(p\, ;l)
k_{\mu_1}' \ldots k_{\mu_l}' \; ,
\label{contracted-projector-boson}
\end{equation}
where
$k$ and $k'$ stand for arbitrary 4-momenta. The inverse relation
\begin{eqnarray}
\label{contracted-projector-inverse}
&& \Pi_{\nu_1 \ldots \nu_l}^{\mu_1 \ldots \mu_l}(p\, ;l) =
\nonumber \\ &&
\frac{1}{(l!)^2}
\der{k^{\nu_1}} \ldots \der{k^{\nu_l}}
\der{k'_{\mu_1}} \ldots \der{k'_{\mu_l}}
{\cal P}^{(l)}(k,k',p) \ \
\end{eqnarray}
gives the form of spin sum
(\ref{spin-sum-boson})
when the explicit expression for contracted projector
(\ref{contracted-projector-boson})
is known%
\footnote{
In the case under consideration we only need to know the form of the
contracted projector with two external momenta ($k_\mu$ and
$k_\mu'$).
}.
As shown in the above-cited articles, this expression reads
\begin{equation}
{\cal P}^{(l)}(k , k' , p) =
\frac{(-1)^l\; l!}{(2l-1)!!}
|\hat{k}|^l |\hat{k'}|^l P_l \left(
\frac{(\hat{k} \hat{k'})}{|\hat{k}| |\hat{k'}| }
\right) \; .
\label{contracted-projector-boson-explicit}
\end{equation}
Here the following compact notations are used:
\begin{eqnarray}
\label{compact-notations-momenta}
&& \hat{k}_\mu \equiv k_\mu -
\frac{(pk)}{M^2}\; p_\mu\; ;\ \ \ \ \ \ \ \
{\hat{k'}}_\mu \equiv k'_\mu -
\frac{(pk')}{M^2}\;  p_\mu\; ;
\nonumber\\&&
|\hat{r}| \equiv \sqrt{|(\hat{r}\hat{r})|}\; ,
(r=k,k').
\end{eqnarray}

Let us now consider the case of a fermion with spin
$J=l+1/2$ ($l=0,1,\ldots$),
mass parameter
$M$
and polarization
$j = -(l + \frac{1}{2}) ,\ldots , +(l + \frac{1}{2})$.
Two corresponding wave functions (particle and anti-particle) are
defined as symmetric traceless spin-tensors%
\footnote{
To save space in this Appendix we use the spinor notations accepted in
\cite{Bogoliubov}.
The correspondence with conventionally used notations (see, e.g.,
\cite{Peskin})
is provided by the relations:
${\cal U}^+ \equiv v$,
${\cal U}^- \equiv u$,
$\overline{{\cal U}}^+ \equiv \overline{u}$,
$\overline{{\cal U}}^- \equiv \overline{v}$.
}
${\cal U}^{\; \pm}_{\rho\mu_1\ldots\mu_l}(j,p)$
and
$\overline{\cal U}^{\; \pm}_{\rho\mu_1\ldots\mu_l}(j,p) \equiv$
$\left( {\cal U}^{\; \mp} \right)^\dagger \gamma_0$.
Here the vector indices
$\mu_1\ldots\mu_l = 0,1,2,3$,
while the spinor one
$\rho = 1,2,3,4$.
The set of auxiliary conditions
(at $p^2 = M^2$)
reads:
\begin{eqnarray*}
&& {\cal U}^{\; \pm}_{\rho\ldots\mu_m\ldots\mu_n\ldots}(j,p) =
{\cal U}^{\; \pm}_{\rho\ldots\mu_n\ldots\mu_m\ldots}(j,p)
\ \ \ \ \text{(symmetry)},
\nonumber\\
\nonumber\\&&
g^{\mu_m\mu_n}
{\cal U}^{\; \pm}_{\rho\ldots\mu_m\ldots\mu_n\ldots}(j,p) = 0
\ \ \ \
\text{(tracelessness)},
\nonumber\\
\nonumber\\&&
p^{\mu_m}
{\cal U}^{\; \pm}_{\rho\ldots\mu_m\ldots}(j,p)=0
\ \ \ \ \text{($p$-transversality)},
\nonumber\\
\nonumber\\&&
\left(\gamma^{\mu_m}\right)_{\rho\tau}
{\cal U}^{\; \pm}_{\tau\ldots\mu_m\ldots}(j,p)=0
\ \ \ \ \text{($\gamma$-transversality)},
\nonumber\\
\nonumber\\&&
\sum\limits_{\rho ;\mu_1 \ldots \mu_l}
\overline{\cal U}^{\; +}_{\rho\mu_1\ldots\mu_l}(i,p)
{\cal U}^{\; -\,\mu_1\ldots\mu_l}_{\rho}(j,p) = (-1)^l\; 2M \delta_{ij}
\nonumber\\&&
\ \ \ \ \ \ \ \ \ \ \ \ \ \ \ \ \ \
\ \ \ \ \ \ \ \ \ \ \ \ \ \ \ \ \ \
\ \ \ \ \ \ \ \ \
\text{(normalization)},
\nonumber\\
\nonumber\\&&
({\rlap/ p} \pm M)_{\rho\tau}
{\cal U}^{\; \pm}_{\tau\mu_1\ldots\mu_l}(j,p) = 0,
\nonumber\\&&
\overline{\cal U}^{\; \pm}_{\rho\mu_1\ldots\mu_l}(j,p)
({\rlap/ p} \mp M)_{\rho\tau} = 0 \ \ \ \
\text{(wave equations)}.
\end{eqnarray*}

In this case the spin sum
\begin{eqnarray}
&& \Pi^{\mu_1\ldots\mu_l}_{\nu_1\ldots\nu_l\;\; \rho \tau}(p\, ;l+1/2)
\equiv
\nonumber \\&&
\sum\limits_{j= -(l + 1/2)}^{j= +(l + 1/2)}
{\cal U}_{\rho \; \nu_1 \ldots \nu_l}^{\; -\;\; }(j,p)
\overline{\cal U}^{\; + \; \mu_1\ldots\mu_l}_{\tau\, }(j,p)
\label{spin-sum-fermion}
\end{eqnarray}
as well as the corresponding contracted projector
\begin{eqnarray}
&&{\cal P}^{(l+\frac{1}{2})}_{\rho\tau}(k, k', p) \equiv \ \ \ \
\nonumber\\&&
k^{\nu_1} \ldots k^{\nu_l}
\Pi^{\mu_1\ldots\mu_l}_{\nu_1\ldots\nu_l\;\; \rho \tau}(p\, ;l+1/2)
{k'}_{\mu_1} \ldots {k'}_{\mu_l}
\label{cpr5}
\end{eqnarray}
are
$4 \times 4$
matrices in spinor space (in what follows we omit spinor
indices). The relation analogous to
(\ref{contracted-projector-boson-explicit})
reads
\begin{widetext}
\begin{equation}
{\cal P}^{(l+\frac{1}{2})}(k,k',p) =
\frac{(-1)^l\; l!}{(2l+1)!!}
|\hat{k}|^l |\hat{k'}|^l
\left\{
{P'}_{l+1} \left(
\frac{\hat{k}_\mu {\hat{k'}}^\mu}{|\hat{k}| |\hat{k'}| }
\right) -
\frac{\s{\hat{k}}\s{\hat{k'}}}{|\hat{k}| |\hat{k'}|}
{P'}_l \left(
\frac{\hat{k}_\mu \hat{k'}^\mu}{|\hat{k}| |\hat{k'}| }
\right)
\right\} (\s{p} + M) \; ,
\label{contracted-projector-fermion-explicit}
\end{equation}
\end{widetext}
where
$P_l (\chi)$
is the Legendre polynomial and
$P_l' (\chi)$
is its derivative with respect to
$\chi$.
Note that at
$p^2 = M^2$
the matrices
$\s{\hat{k}} \s{\hat{k'}}$
and
$\s{p}$
commute with one another, so the position of the term
$(\s{p} + M)$
does not matter as long as one only needs to know the numerator of
minimal propagator.

\section{Resonance exchange contributions}
\label{app-graphs}
\mbox

Below we give the full list of contributions to invariant amplitudes
$A^{\pm}$
and
$B^{\pm}$
provided by the resultant graphs with
$s$-
and
$t$-channel
exchanges (see
Fig.~\ref{1f}).
The contributions of the graphs with
$u$-channel
resonances can be obtained from those with
$s$-channel
ones with the help of corresponding substitution.

We employ the notations of
Sec.~\ref{sec-General};
the auxiliary functions
$
\Phi (M,m,\mu),
$
$
F_A^{l}(M,\chi)
$
and
$
F_B^{l}(M,\chi)
$
are defined in
(\ref{Kallen-function-Phi}) ---
(\ref{function-FBl}).
As always throughout the paper,
$m$
and
$\mu$
are the nucleon and pion masses,
$M$
is the resonance mass,
$P_l(\chi)$
is the Legendre polynomial and
$P'_l(\chi)$ ---
its derivative with respect to argument. Introducing the abbreviation
$(\s{k}+\s{k}')/2 \equiv \s{Q}$
and compact notation for baryon numerical factors:
\begin{equation}
G(\pi N B) = \left| g_{\sss B} (J,I,{\cal N}_B) \right|^2
(\Phi)^l
\frac{l!}{(2l+1)!!} \; ,
\label{G-pi-N-baryon}
\end{equation}
where
$B$
refers to any relevant baryon of spin
$J= l + 1/2$,
isospin
$I$
and normality
${\cal N}_B$,
we get the following expressions for the
$x$-channel
$(x = s,u)$
baryon exchange graphs (see
Fig.~\ref{gr1f} {\bf a, b}):

\begin{figure*}
\includegraphics[height=3cm]{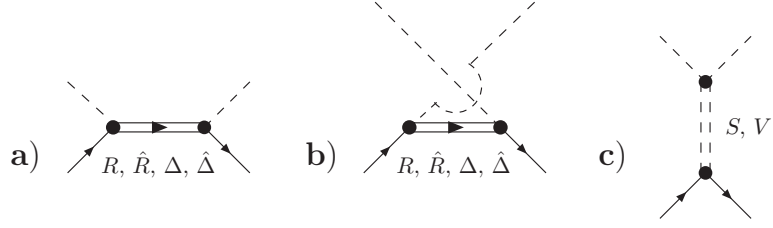}
\caption{\label{gr1f}
Resonance exchange graphs:
{\bf a:}
$I=1/2, \,3/2$ in
$s$-channel;
\ \
{\bf b:}
$I=1/2, \,3/2$ in
$u$-channel;
\ \
{\bf c:}
$I=0, \,1$ in
$t$-channel; }
\end{figure*}

\begin{widetext}
\begin{equation}
T^{I}_{\beta\alpha, ba}(Z_x)\,
\overline{u} (p',\lambda')
\left\{
-\, \frac{G(\pi N B)}{x - M_{\sss R}^2}
\left[
F^l_A (-{\cal N}_B M,t) + Z_x\, \s{Q} F^l_B (-{\cal N}_B M, t)
\right] \right\}
u(p,\lambda)\, .
\end{equation}
\end{widetext}
Here
$$
Z_s = 1,\ \ \ \ Z_u = -1,
$$
and
\begin{eqnarray*}
&& T^{1/2}_{\beta\alpha, ba}(Z_x) \equiv
\left( \delta_{ba}\delta_{\beta\alpha} +
Z_x\, i\varepsilon_{bac} (\sigma_c)_{\beta\alpha} \right);
\nonumber\\ &&
T^{3/2}_{\beta\alpha, ba}(Z_x) \equiv
\left( -\, \frac{2}{3} \delta_{ba}\delta_{\beta\alpha} +
Z_x\, \frac{i}{3}\, \varepsilon_{bac}(\sigma_c)_{\beta\alpha} \right).
\end{eqnarray*}

Further, using
\begin{equation}
F(M,m,\mu) \equiv \frac{1}{4}\sqrt{|(M^2-4m^2)(M^2-4\mu^2)|}
\label{Kallen-function-F}
\end{equation}
and the compact notation for meson numerical factors
($I=S,V$
stands for isoscalar and isovector, respectively; the constants
$g{\sss I \pi \pi}$
and
$g_{\sss NNI}^{(n)}$
are introduced in
(\ref{H-Spipi}) -- (\ref{H-VNN}))
\begin{equation}
G_{n}^I \equiv (F)^J g{\sss I \pi \pi}\, g_{\sss NNI}^{(n)}\,
\frac{J!}{(2J-1)!!}\, ,\ \ \ \ \ \
(n=1,2)\; ,
\label{G-pipi-NN}
\end{equation}
we obtain the contributions due to the meson exchange graphs
(see
Fig.~\ref{gr1f} {\bf c}):
$$
T^I_{ba,\beta\alpha}\,
\overline{u} (p',\lambda')\,
{\cal M}^{I}\, u(p,\lambda)\, ,
$$
where
$$
T^S_{ba,\beta\alpha}\, \equiv -\,\delta_{ba}\; \delta_{\beta\alpha}\, ,\ \ \ \ \ \ \
T^V_{ba,\beta\alpha}\, \equiv i \varepsilon_{bac} (\sigma_c)_{\beta\alpha}\, ,
$$
and
\begin{widetext}
$$
{\cal M}^{I} \equiv
\frac{G_1^I}{t - M_{\sss I}^2}P_{\sss J} \left( \frac{s - u}{4F} \right) +
\frac{G_2^I}{t - M_{\sss I}^2}
\left\{
\frac{4m}{M_{\sss I}^2 - 4m^2}
P'_{\sss J-1} \left( \frac{s - u}{4F} \right) +
\s{Q} \frac{1}{F} P'_{\sss J}
\left( \frac{s - u}{4F} \right)
\right\}
$$
\end{widetext}
(it is assumed that
$P'_{\sss -1}(\chi)\equiv 0$).
Note that in this expression
$t$
and
$s-u \equiv \nu_t$
are independent variables; the residue at
$t = M^2$
may (and does) depend on
$\nu_t$.


\section{List of baryon resonances}
\label{app-resonance-list}

In Table \ref{BaryonDat} we
give a list of
$N$
and
$\Delta$
baryon resonances which we take into account when testing our sum
rules. Here
$J$, $I$, $P$
and
${\cal N}$
stand for the resonance spin, isospin, parity and normality,
respectively. The minimal and maximal values of
$G(\pi NR)$
have been calculated using the data from
\cite{PDG}.

\begin{table}[H]
\caption{\label{BaryonDat} $N$ and $\Delta$-baryon summary table.}
\begin{ruledtabular}
\begin{tabular}{lcccl}
$\; \; R$ & $I$ & $J$ & $P({\cal N})\ \ $& $G(\pi NR)$ \\
\hline
$N(940)$  &  $1/2$& $1/2$ &$+(+)$ & $179.\,7$ \\
$N(1440)$  & $1/2$ & $1/2$ & $+(+)$ & $26.\,0 \div 63.\,9$ \\
$N(1680)$  & $1/2$ & $5/2$ & $+(+)$ & $6.\,06 \div 7.\,60 $\\
$N(1710)$  & $1/2$&  $1/2$ &  $+(+)$ & $0.\,36 \div 3.\,6 $\\
$N(1720)$  & $1/2$ & $3/2$ & $+(-)$ & $0.\,09 \div 0.\,35 $\\
$N(2220)$  & $1/2$ & $9/2$ & $+(+)$ & $ 0.\,94 \div 2.\,7 $\\
$N(1520)$ &  $1/2$ & $3/2$ & $-(+)$ & $7.\,35 \div 10.\,9$ \\
$N(1535)$ &  $1/2$ & $1/2$ & $ -(-) $& $0.\,30 \div 0.\,67$ \\
$N(1650)$ &  $1/2$ & $1/2$ & $-(-)$ & $0.\,54 \div 1.\,09$ \\
$N(1675)$ &  $1/2$ & $5/2$ & $ -(-)$ & $0.\,28 \div 0.\,45 $\\
$N(1700)$ &  $1/2$ & $3/2$ & $ -(+) $& $0.\,19 \div 1.\,68 $\\
$N(2190)$ &  $1/2$ & $7/2$ & $ -(+)$ & $0.\,9 \div 4.\,3 $\\
$N(2250)$ &  $1/2$ & $9/2$ & $-(-)$ & $0.\,05 \div 0.\,53$ \\
$N(2600)$ &  $1/2$ & $11/2$& $-(+)$ & $0.\,43 \div 1.\,40$ \\
\hline
\hline
$\; \; R$ & $I$ & $J$ & $P({\cal N})\ \ $& $G(\pi NR)$ \\
\hline
$\Delta (1232)$ & $3/2$ & $3/2$ & $+(-)$ & $4.\,17 \div 4.31 $\\
$\Delta (1600)$ & $3/2$ & $3/2$ & $+(-)$ & $0.\,49 \pm 2.\,2 $\\
$\Delta (1905)$ & $3/2$ & $5/2$ & $+(+)$ & $3.5 \div 8.7 $\\
$\Delta (1910)$ & $3/2$ & $1/2$ & $+(+)$ & $4.\,8 \div 9.\,7 $\\
$\Delta (1920)$ & $3/2$ & $3/2$ & $+(-)$ & $0.\,12 \div 0.\,94$ \\
$\Delta (1950)$ & $3/2$ & $7/2$ & $+(-)$ & $1.\,29 \div 3.\,36$ \\
$\Delta (2420)$ & $3/2$ & $11/2$& $ +(-)$ &$ 0.\,19 \div 0.\,97$ \\
$\Delta (1620)$ & $3/2$ & $1/2$ & $-(-)$ & $0.\,51 \div 0.\,86$  \\
$\Delta (1700)$ & $3/2$ & $3/2$ & $-(+)$ & $4.\,5 \div 17.\,9 $\\
$\Delta (1930)$ & $3/2$ & $5/2$ & $-(-)$ & $0.\,19 \div 1.\,04 $\\
$\Delta (1950)$ & $3/2$ & $7/2$ & $+(-)$ & $1.\,3 \div 2.\,4 $\\
$\Delta (2420)$ & $3/2$ & $11/2$ & $+(-)$ & $ 0.\,2 \div 1.\,0$\\
\end{tabular}
\end{ruledtabular}
\end{table}


\end{document}